\documentclass[superscriptaddress,aps,pra,twocolumn,showpacs,preprintnumbers,amsmath,amssymb,floatfix,groupedaddress]{revtex4}

\usepackage[dvips]{graphicx}
\usepackage{dcolumn}
\usepackage{bm}
\usepackage{amssymb}
\usepackage{xr}

\newcommand{\eop}{\mathcal{E}}

\begin{document}

\title{Photon storage in $\Lambda$-type optically dense atomic media. II. Free-space model}

\author{Alexey V. Gorshkov}
\author{Axel Andr\'e}
\author{Mikhail D. Lukin}
\affiliation{Physics Department, Harvard University, Cambridge, Massachusetts 02138, USA}
\author{Anders S. S{\o}rensen}
\affiliation{QUANTOP, Danish National Research Foundation Centre of
Quantum Optics, Niels Bohr Institute, DK-2100 Copenhagen {\O},
Denmark}

\date{\today}


\begin{abstract}
In a recent paper [Gorshkov \textit{et al.}, Phys. Rev. Lett. \textbf{98}, 123601 (2007)], we presented a universal physical picture for describing a wide range of techniques for storage and retrieval of photon wave packets in $\Lambda$-type atomic media in free space, including the adiabatic reduction of the photon group velocity, pulse-propagation control via off-resonant Raman techniques, and photon-echo based techniques. This universal picture produced an optimal control strategy for photon storage and retrieval applicable to all approaches and yielded identical maximum efficiencies for all of them. In the present paper, we present the full details of this analysis as well some of its extensions, including the discussion of the effects of non-degeneracy of the two lower levels of the $\Lambda$ system.
The analysis in the present paper is based on the intuition obtained from the study of photon storage in the cavity model in the preceding paper [Gorshkov \textit{et al.}, Phys. Rev. A \textbf{76}, 033804 (2007)].
\end{abstract} 



\maketitle

\section{Introduction}

High fidelity storage of a traveling pulse of light into an atomic memory and the subsequent retrieval of the state back onto a light pulse are currently being pursued by a number of laboratories around the world. In a recent paper \cite{gorshkov07}, we presented a universal picture for describing, optimizing, and showing a certain degree of equivalence between a wide range of techniques for photon storage and retrieval in $\Lambda$-type atomic media, including the approaches based on electromagnetically induced transparency (EIT), off-resonant Raman processes, and photon echo. In the present paper, as well as in the preceding paper \cite{paperI} (which we will refer to henceforth as paper I) and the paper that follows \cite{paperIII} (which we will refer to henceforth as paper III), we present all the details behind this universal picture and the optimal control shaping that it implies, as well as consider several extensions of this analysis beyond the results of Ref.~\cite{gorshkov07}. In particular, in paper I, we present the full details of the optimization in the slightly simpler model where the atoms are placed inside a cavity. Using the intuition gained from the cavity model discussion, we show in the present paper all the details behind the analysis of the free-space model given in Ref.~\cite{gorshkov07}. We also discuss several extensions of the analysis of Ref.~\cite{gorshkov07}, such as the inclusion of the decay of coherence between the two lower levels of the $\Lambda$ system and the effects of nondegeneracy of these two levels. Finally, in paper III, we generalize our treatment to include the effects of inhomogeneous broadening.

For a complete introduction to photon storage in $\Lambda$-type atomic media, as it applies to paper I and to the present paper, as well as for the full list of references, we refer the reader to paper I. In the present Introduction, we only list the two main results of the present paper. The first important result is the abovementioned proof of a certain degree of equivalence between a variety of different photon storage protocols. In particular, this result means that provided there is a sufficient degree of control over the shape of the incoming photon wave packet and/or over the power and shape of the classical control pulses, all the protocols considered have the same maximum achievable efficiency that depends only on the optical depth $d$ of the medium. The second important result is a novel time-reversal-based iterative algorithm for optimizing quantum state mappings, a procedure that we expect to be applicable beyond the field of photon storage. One of the key features of this optimization algorithm is that it can not only be used as a mathematical tool but also as an experimental technique. In fact, following our theoretical proposal, an experimental demonstration of this technique has already been carried out \cite{novikova07}. Both the experimental results \cite{novikova07} and the theoretical results of the present paper indicate that the suggested optimization with respect to the shape of the incoming photon wave packet and/or the control pulse shape and power will be important for increasing the photon storage efficiencies in current experiments.

Although the slightly simpler cavity model discussed in paper I is similar enough to the free-space model to provide good intuition for it, the two physical systems have their own advantages and disadvantages, which we will discuss in the present paper. One advantage of the free-space model is the fact that it is easier to set up experimentally, which is one of the reasons we study this model in the present paper. Turning to the physics of the two models, the main differences come from the fact that in the cavity model the only spin wave mode accessible is the one that has the excitation distributed uniformly over all the atoms. In contrast, in the free-space model, incoming light can couple to any mode specified by a smooth excitation with position-dependent amplitude and phase. As a consequence of this, the free-space model allows for high efficiency storage of a wider range of input light modes than the cavity model. In particular, we showed in paper I that in the cavity model high efficiency photon-echo-based storage (which we refer to as fast storage) is possible for a single input mode of duration $\sim 1/(\gamma C)$, where $\gamma$ is the optical polarization decay and $C$ is the cavity cooperativity parameter. In contrast, we show in the present paper that high efficiency fast storage in a free-space atomic ensemble with optical depth $d$ is possible for any input light mode of duration $T$ provided $T \gamma \ll 1$ and $T d \gamma \gg 1$. However, the cavity model also has some advantages over the free-space model. In particular, the error during optimal light storage and retrieval for a given atomic ensemble scales as the inverse of the optical depth, as we have shown for the cavity model in paper II and for the free-space model in the present paper. The optimal efficiency is therefore higher when the ensemble is enclosed in a cavity, which effectively enhances the optical depth by the cavity finesse to form the cooperativity parameter $C$. Moreover, if one is forced to retrieve from a spatially uniform spin wave mode (e.g., if the spin wave is generated via spontaneous Raman scattering \cite{eisaman05}), the error during retrieval will decrease faster with optical depth in the cavity model ($\sim 1/C$) than in the free-space model ($\sim 1/\sqrt{d}$).   

The remainder of the present paper is organized as follows. In Sec.~\ref{sec:freespacemodel}, the model is introduced. In Secs.~\ref{sec:freeret}, \ref{sec:timerev}, and \ref{sec:timerev2}, we prove that during retrieval there exists a fixed branching ratio between the desired light emission rate and undesired polarization decay rate, and use this in combination with time reversal to derive the optimal strategy for storage and retrieval without fully solving the equations. In Secs.~\ref {sec:freead} and \ref{sec:freefast}, the equations are solved analytically in the adiabatic and fast limits, respectively, and more specific statements about the optimal control strategy are made. In Sec.~\ref{sec:nondeg}, the effect of nondegeneracy of the two metastable states is discussed. In Sec.~\ref{sec:freesum}, we summarize the discussion of the free-space model. Finally, in the Appendixes, we present some details omitted in the main text. 

\section{Model \label{sec:freespacemodel}}

We refer the reader to Appendix \ref{sec:appModelfree} for the details of the model and for the derivation of the equations of motion. In this section, we only briefly summarize the model and state the equations of motion without derivation.

We consider a free-space medium of length $L$ and cross-section area $A$ containing $N = \int_0^L d z n(z)$ atoms, where $n(z)$ is the number of atoms per unit length. We assume that within the interaction volume the concentration of atoms is uniform in the transverse direction. The atoms have the same $\Lambda$-type level configuration as in the cavity case discussed in paper I and shown in Fig.~1 of paper I. They are coupled to a quantum field and a copropagating classical field. We assume that quantum electromagnetic field modes with a single transverse profile are excited. We also assume that both the quantum and the classical field are narrowband fields centered at $\omega_1 = \omega_{eg} - \Delta$ and $\omega_2 = \omega_{es}-\Delta$, respectively (where $\omega_{eg}$ and $\omega_{es}$ are atomic transition frequencies). The quantum field is described by a slowly varying operator $\hat \eop(z,t)$, while the classical field is described by the Rabi frequency envelope $\Omega(z,t) = \Omega(t-z/c)$.  

We neglect reabsorption of spontaneously emitted photons. This is a good approximation since we are interested in the storage of single- or few-photon pulses, in which case there will be at most a few spontaneously emitted photons. Although for an optically thick medium they can be reabsorbed and reemitted \cite{fleischhauer99, matsko01}, the probability of spontaneously emitting into the mode $\hat \eop$ is given by the corresponding far-field solid angle $\sim \lambda^2/A \sim d/N$, where $A$ is the cross section area of both the quantum field mode and the atomic medium (see Appendix \ref{sec:appModelfree} for a discussion of why this choice is not important), $\lambda = 2 \pi c/\omega_1$ is the wavelength of the quantum field, and $d \sim \lambda^2 N/A$ is the resonant optical depth of the ensemble. In most experiments, this probability is very small. Moreover, we will show that for the optimized storage process, the fraction of the incoming photons lost to spontaneous emission will decrease with increasing optical depth. In practice, however, reabsorption of spontaneously emitted photons
can cause problems \cite{fleischhauer99b} during the optical pumping process, which is used to initialize the sample, and this may require modification of the present model.

We treat the problem in a one-dimensional approximation. This is a good approximation provided that the control beam is much wider than the single mode of the quantum field defined by the optics, as, for example, in the experiment of Ref.~\cite{eisaman05}. In this case, the transverse profile of the control field can be considered constant; and, in the paraxial approximation, the equations reduce to one-dimensional equations for a single Hermite-Gaussian quantum field mode \cite{andre05, sorensen07}.

We define the polarization operator $\hat P(z,t) = \sqrt{N} \hat \sigma_{ge}(z,t)$ and the spin-wave operator $\hat S(z,t) = \sqrt{N} \hat \sigma_{gs}(z,t)$ (where $\hat \sigma_{\mu \nu}(z,t)$ are slowly varying position-dependent collective atomic operators defined in Appendix \ref{sec:appModelfree}). In the dipole and rotating-wave approximations, to first order in $\hat \eop$, and assuming that at all times almost all atoms are in the ground state, the Heisenberg equations of motion read
\begin{eqnarray} \label{freeeqse}
\!\!\!\!\!\!\!\!\!\!\!(\partial_t + c \partial_z) \hat \eop \!\! &=& \!\! i g \sqrt{N} \hat P n(z) L/N,
\\
\partial_t \hat P \!\!&=&\!\! - (\gamma+i \Delta) \hat P \!+\! i g \sqrt{N} \hat \eop \!+\! i \Omega \hat S \!+\! \sqrt{2 \gamma} \hat F_P,
\\ \label{freeeqss}
\partial_t \hat S \!\!&=&\!\! -\gamma_\textrm{s} \hat S + i \Omega^* \hat P + \sqrt{2 \gamma_s} \hat F_S,
\end{eqnarray}
where we introduced the spin-wave decay rate $\gamma_\textrm{s}$, the polarization decay rate $\gamma$, and the corresponding Langevin noise operators $\hat F_P(z,t)$ and $\hat F_S(z,t)$. As in the cavity case, collective enhancement \cite{lukin03} results in the increase of the atom-field coupling constant $g$ (assumed to be real for simplicity) by a factor of $\sqrt{N}$ up to $g \sqrt{N}$.

As we show in Appendix \ref{sec:appModelfree} and explain in detail in paper I, under reasonable experimental conditions, the normally ordered noise correlations of $\hat F_P$ and $\hat F_S$ are zero, i.e., the incoming noise is vacuum and the transformation is passive. As we show in Sec.~II of paper I, this implies that efficiency is the only number required to completely characterize the mapping.

As in the cavity discussion of paper I, we suppose that initially all atoms are in the ground state, i.e.,  no atomic excitations are present. We also assume that there is only one nonempty mode of the incoming quantum field and that it has an envelope shape $h_0(t)$ nonzero on $[0,T]$. The term ``photon storage and retrieval" refers to mapping this mode onto some mode of $\hat S$ and, starting at a later time $T_\textrm{r} > T$, retrieving it onto an outgoing field mode. Then precisely as in the cavity case in paper I, for the purposes of finding the storage efficiency, which is given by the ratio of the numbered of stored excitations to the number of incoming photons
\begin{equation}
\eta_{\textrm{s}} = \frac{\int_0^L d z \frac{n(z)}{N} \langle \hat S^\dagger(z,T) \hat S(z,T) \rangle}{\frac{c}{L} \int_0^T  d t  \langle \hat \eop^\dagger(0,t) \hat \eop(0,t) \rangle},
\end{equation}
we can ignore $\hat F_P$ and $\hat F_S$ in Eqs.~(\ref{freeeqse})-(\ref{freeeqss}) and treat these equations as complex number equations with the interpretation that the complex number fields describe the shapes of quantum modes. In fact, although the resulting equations describe our case of quantized light coupled to the $|g\rangle-|e\rangle$ transition, they will also precisely be the equations describing the propagation of a classical probe pulse. To see this, one can simply take the expectation values of Eqs.~(\ref{freeeqse})-(\ref{freeeqss}) and use the fact that classical probe pulses are described by coherent states.  

Going into a comoving frame $t' = t - z/c$, introducing dimensionless time $\tilde t = \gamma t'$ and dimensionless rescaled coordinate $\tilde z = \int_0^z d z' n(z')/N$, absorbing a factor of $\sqrt{c/(L \gamma)}$ into the definition of $\eop$, we reduce Eqs.~(\ref{freeeqse})-(\ref{freeeqss}) to
\begin{eqnarray} \label{freeeqs2e}
\partial_{\tilde z} \eop &=& i \sqrt{d} P,
\\ \label{freeeqs2p}
\partial_{\tilde t} P &=& - (1 + i \tilde \Delta) P + i \sqrt{d} \eop + i \tilde \Omega (\tilde t) S,
\\ \label{freeeqs2s}
\partial_{\tilde t}  S &=& i \tilde \Omega^*(\tilde t) P,
\end{eqnarray}  
where we have identified the optical depth $d = g^2 N L/(\gamma c)$ and where $\tilde \Delta = \Delta/\gamma$ and $\tilde \Omega = \Omega/\gamma$. We confirm in Appendix \ref{sec:appModelfree} that from the definition it follows that $d$ is independent of the size of the beam and, for a given transition, only depends on the density and length of the ensemble. Moreover, the definition of $d$ that we use here can be related to the intensity attenuation of a resonant probe in our three level system with the control off, in which case the equations give an attenuation of $\exp(-2 d)$. In Eqs.~(\ref{freeeqs2e})-(\ref{freeeqs2s}) and in the rest of this paper (except for Sec.~\ref{sec:freespindecay}), we neglect the decay $\gamma_\textrm{s}$ of the spin wave. However, precisely as in the cavity case, nonzero $\gamma_\textrm{s}$ simply introduces an exponential decay without making the solution or the optimization harder, as we will discuss in Sec.~\ref{sec:freespindecay}. We also note that Eqs.~(\ref{freeeqse})-(\ref{freeeqss}) are the same as the equations of Ref.~\cite{fleischhauer02} for copropagating fields, generalized to nonzero $\Delta$ and $\gamma_\textrm{s}$, and taken to first order in $\eop$.

\begin{figure}[ht]
\includegraphics[scale = 0.35]{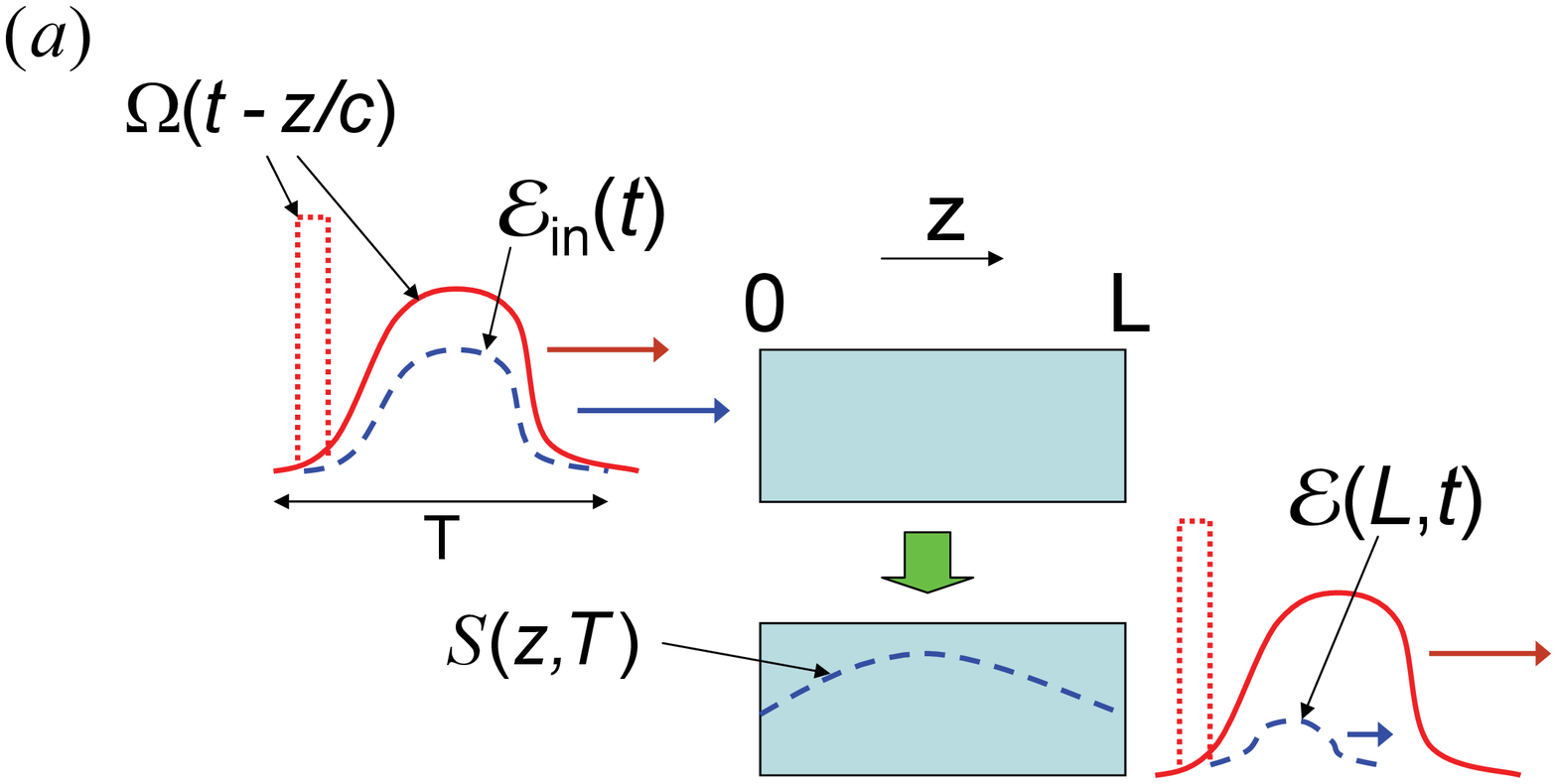}
\includegraphics[scale = 0.35]{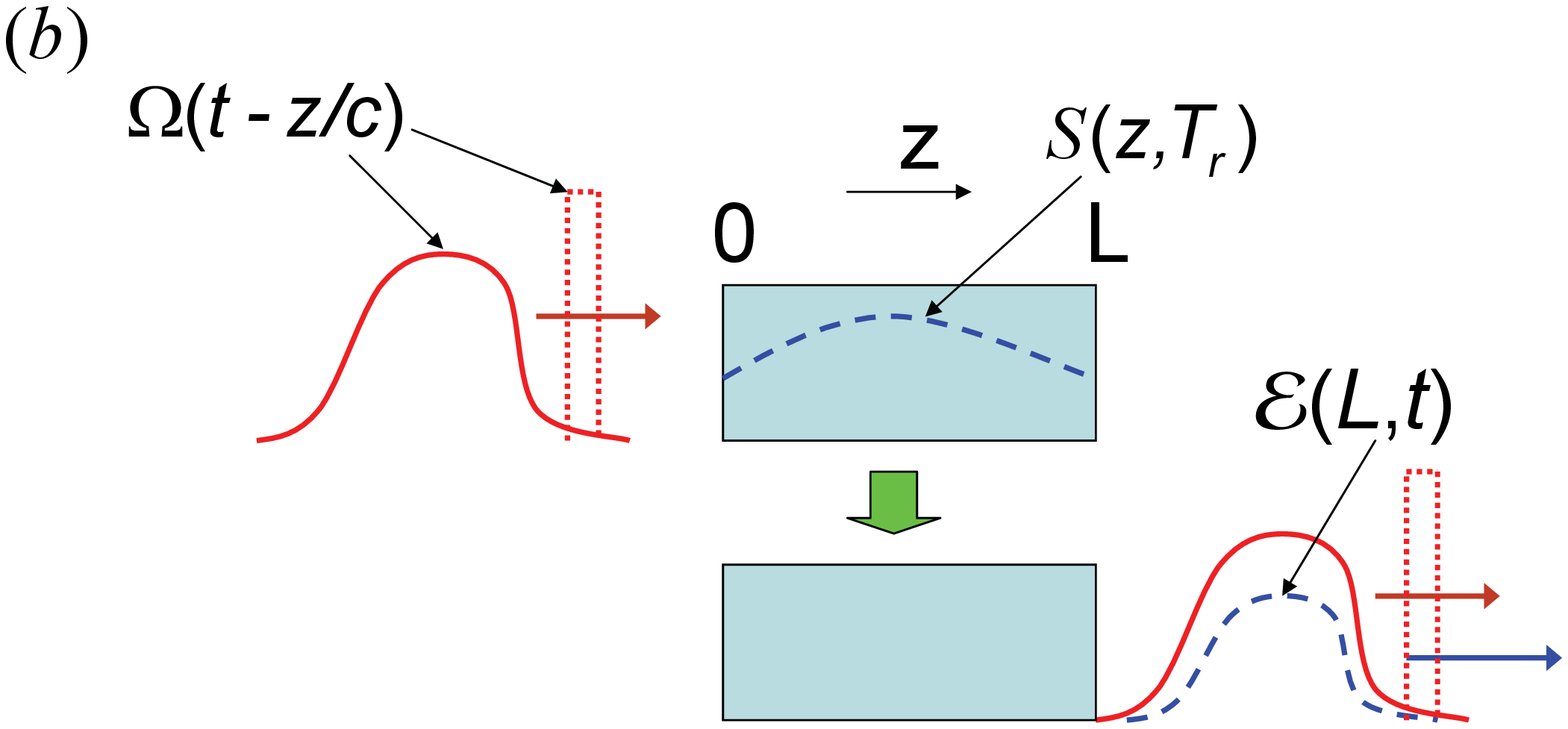}
\includegraphics[scale = 0.35]{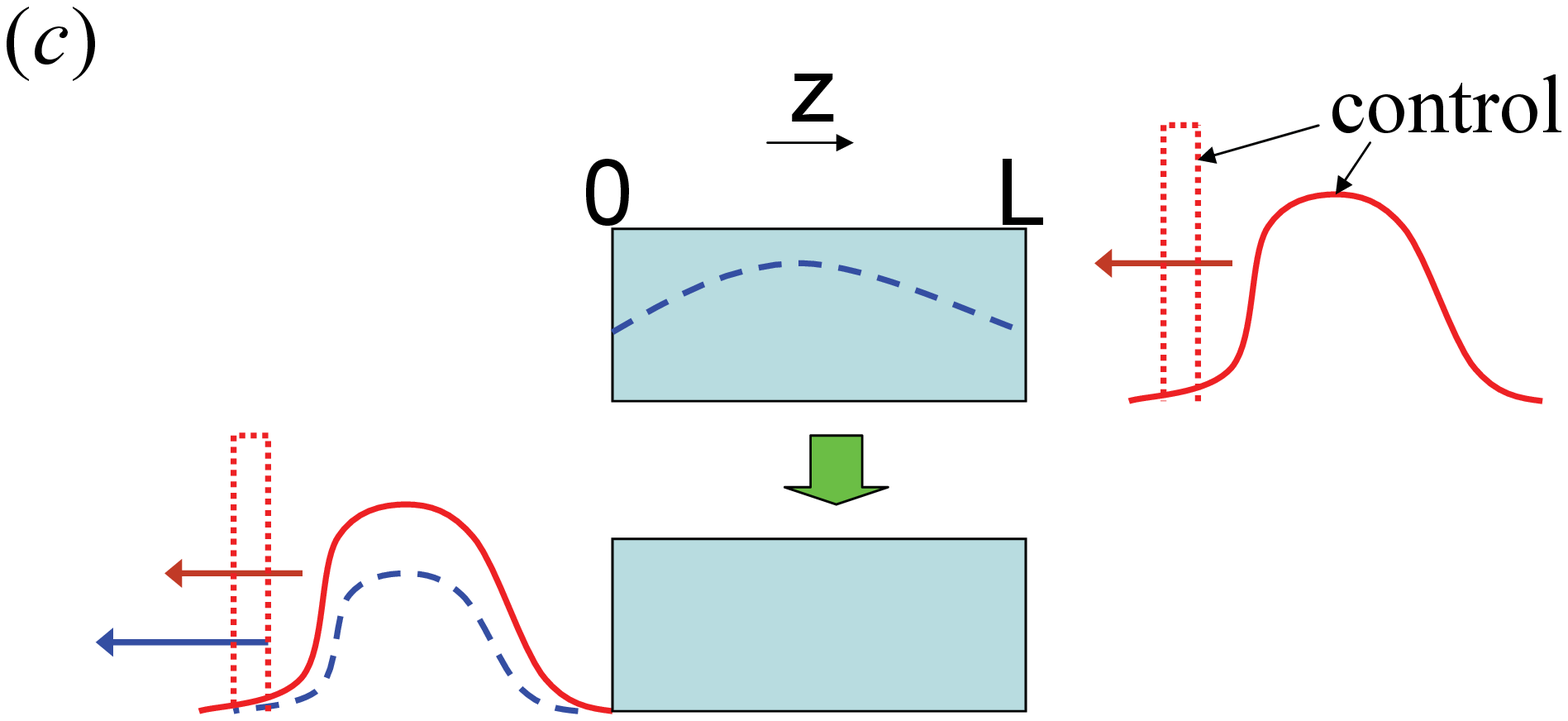}
\caption{(Color online) (a) Storage, (b) forward retrieval, and (c) backward retrieval setup. The smooth solid  curve is the generic control field shape ($\Omega$) for adiabatic storage or retrieval; the dotted square pulse indicates a $\pi$-pulse control field for fast storage or retrieval. The dashed line indicates the quantum field $\eop$ and the spin-wave mode $S$.  During storage, $\eop(L,t)$ is the ``leak," whereas it is the retrieved field during retrieval. \label{fig:setup}}  
\end{figure}

During storage, shown (in original variables) in Fig.~\ref{fig:setup}(a), the initial and boundary conditions are (in rescaled variables)  $\eop(\tilde z = 0, \tilde t) = \eop_\textrm{in}(\tilde t)$, $P(\tilde z,\tilde t=0) = 0$, and $S(\tilde z,\tilde t=0) = 0$, where $\eop_\textrm{in}(\tilde t)$ is nonzero for $\tilde t \in [0,\tilde T]$ (where $\tilde T = T \gamma$) and, being a shape of a mode, is normalized according to $\int_0^{\tilde T} d \tilde t |\eop_\textrm{in}(\tilde t)|^2 = 1$. $S(\tilde z, \tilde T)$ gives the shape of the spin-wave mode, into which we store, and the storage efficiency is given by
\begin{equation}
\eta_\textrm{s} = \int_0^1 d \tilde z |S(\tilde z, \tilde T)|^2.
\end{equation}
Loss during storage comes from the decay $\gamma$ as well as from the ``leak" $\eop(\tilde z = 1,\tilde t)$ shown in Fig.~\ref{fig:setup}(a). Then at a later time $\tilde T_\textrm{r} > \tilde T$ (where $\tilde T_r = T_r \gamma$), we want to retrieve the excitation back onto a photonic mode either in the forward direction, as shown in Fig.~\ref{fig:setup}(b), or in the backward direction \cite{matsko01b} (i.e., with the retrieval control pulse incident from the right) as shown in Fig.~\ref{fig:setup}(c). Instead of turning our Eqs.~(\ref{freeeqs2e})-(\ref{freeeqs2s}) around to describe backward retrieval, we invert, for backward retrieval, the spin wave according to $S(\tilde z, \tilde T_\textrm{r}) = S(1-\tilde z, \tilde T)$, whereas we keep $S(\tilde z, \tilde T_\textrm{r}) = S(\tilde z, \tilde T)$ for forward retrieval. 
Because of the $z$-dependent phases in Eq.~(\ref{slowops4}), this prescription for backward retrieval is strictly valid only for zero splitting between the two metastable states ($\omega_{sg}=0$). In Sec.~\ref{sec:nondeg}, we will discuss the effect of nonzero $\omega_{sg}$.
The remaining initial and boundary conditions during retrieval are $\eop(\tilde z = 0, \tilde t) = 0$ and $P(\tilde z, \tilde T_\textrm{r}) = 0$. If we renormalize the spin wave before doing the retrieval, then the retrieval efficiency will be given by
\begin{equation} \label{etar}
\eta_\textrm{r} = \int_{\tilde T_\textrm{r}}^\infty d \tilde t |\eop(1,\tilde t)|^2.
\end{equation}
If we do not renormalize the spin wave before doing the retrieval, this formula will give the total efficiency of storage followed by retrieval $\eta_\textrm{tot} = \eta_\textrm{s} \eta_\textrm{r}$.

To solve Eqs.~(\ref{freeeqs2e})-(\ref{freeeqs2s}), it is convenient to Laplace transform them in space according to $\tilde z \rightarrow u$, so that Eqs.~(\ref{freeeqs2e},\ref{freeeqs2p}) become 
\begin{eqnarray} \label{freeeqs2el}
\eop &=& i \frac{\sqrt{d}}{u} P + \frac{\eop_\textrm{in}}{u},
\\ \label{freeeqs2pl}
\partial_{\tilde t} P &=& - (1 +\frac{d}{u} + i \tilde \Delta) P + i \tilde \Omega(t) S + i \frac{\sqrt{d}}{u} \eop_\textrm{in}.
\end{eqnarray}
As in the cavity case in paper I, it is also convenient to reduce Eqs.~(\ref{freeeqs2e})-(\ref{freeeqs2s}) to a single equation 
\begin{equation} \label{freeoneeq}
\left[ \ddot S - \frac{\dot{\tilde{\Omega}}^*}{\tilde \Omega^*} \dot S\right] + (1+\frac{d}{u} + i \tilde \Delta) \dot S + |\tilde \Omega|^2 S = 
-\tilde \Omega^* \frac{\sqrt{d}}{u} \eop_\textrm{in},
\end{equation}
where the overdot stands for the $\tilde t$ derivative. As in the cavity case, this second-order differential equation cannot, in general, be fully solved analytically. Similar to paper I, we can, however,  derive several important results regarding the optimal control strategy for storage and retrieval without making any more approximations. We discuss these results in Sec.~\ref{sec:freeret}, where we optimize retrieval, and in Secs.~\ref{sec:timerev} and \ref{sec:timerev2} where we introduce the important time reversal ideas, which allow us to deduce the optimal storage from the optimal retrieval.

\section{Optimal Retrieval \label{sec:freeret}}

Although Eq.~(\ref{freeoneeq}) cannot, in general, be fully solved analytically, we still make in this and in the following two sections several important statements regarding the optimal strategy for maximizing the storage efficiency, the retrieval efficiency, and the combined (storage followed by retrieval) efficiency without making any more approximations. It is convenient to first consider retrieval, and we do so in this section. 

Although we cannot, in general, analytically solve for the output field $\eop_\textrm{out}(t)$, we will show now that, as in the cavity case in paper I, the retrieval efficiency is independent of the control shape and the detuning provided no excitations are left in the atoms. Moreover, we will show that the retrieval efficiency is given by a simple formula that depends only on the optical depth and the spin-wave mode. From Eqs.~(\ref{freeeqs2s}) and (\ref{freeeqs2pl}), it follows that 
\begin{eqnarray} \label{ddtPPstar}
\frac{d}{d \tilde t}\left(P(u, \tilde t) \left[P(u'^*,\tilde t)\right]^* + S(u, \tilde t) \left[S(u'^*, \tilde t)\right]^*\right) 
\nonumber \\
= - (2+d/u+d/u') P(u,\tilde t) \left[P(u'^*,\tilde t)\right]^*. 
\end{eqnarray}
Using Eqs.~(\ref{freeeqs2el}) and (\ref{ddtPPstar}) and assuming $P(u,\infty) = S(u,\infty)=0$ (i.e., that no excitations are left in the atoms at $\tilde t = \infty$), the retrieval efficiency is
\begin{eqnarray}
\eta_{\textrm{r}} &=& \!\!\mathcal{L}^{-1}\!\left\{\!\frac{d}{u u'} \int_{\tilde T_{\textrm{r}}}^\infty d \tilde t P(u,\tilde t) \left[P(u'^*,\tilde t)\right]^*\right\}
\nonumber \\
&=& 
\!\!\mathcal{L}^{-1}\!\left\{\!\frac{d}{2 u u' + d(u + u')}  S(u, \tilde T_\textrm{r}) \left[S(u'^*,\tilde T_\textrm{r})\right]^*\right\}
\nonumber \\ \label{etargen}
&=&\!\!
\int_0^1 \!\!\!d \tilde z\! \int_0^1 \!\!\!d \tilde z' S(1-\tilde z,\tilde T_{\textrm{r}}) S^*(1-\tilde z',\tilde T_{\textrm{r}}) k_\textrm{r}(\tilde z, \tilde z'),  
\end{eqnarray}
where $\mathcal{L}^{-1}$ means that two inverse Laplace transforms ($u \rightarrow \tilde z$ and $u' \rightarrow \tilde z'$) are taken and are both evaluated at $\tilde z = \tilde z' = 1$ and where the kernel $k_\textrm{r}$ is defined as
\begin{eqnarray} \label{retkernel}
k_\textrm{r}(\tilde z,\tilde z')\!\!\! &=&\!\!\! \mathcal{L}^{-1}\!\left\{\!\frac{d}{2 u u' + d(u + u')} e^{-u (1-\tilde z) - u' (1-\tilde z')} \!\right\}
\nonumber \\
&=&\!\!\! \frac{d}{2} e^{- d \frac{\tilde z'+\tilde z}{2}} I_0(d \sqrt{\tilde z \tilde z'}),
\end{eqnarray}
where $I_n$ is the $n$th-order modified Bessel function of the first kind. Similarly to the cavity case in paper I, we see that the efficiency is independent of $\Delta$ and $\Omega$, which reflects that  in Eq.~(\ref{freeeqs2pl}) (or, equivalently, on the right-hand side of Eq.~(\ref{ddtPPstar}))   there is a fixed  branching ratio between the decay rates of $P$.  For a given $u$ the rates are (in the original units) $\gamma$ and $\gamma d/u$ into the undesired modes and the desired mode $\eop_\textrm{out}$, respectively, independent of $\Delta$ and $\Omega$. In fact, a stronger result than the independence of retrieval efficiency from $\Delta$ and $\Omega$ can be obtained: as we show in Appendix \ref{sec:position}, the distribution of spontaneous emission loss as a function of position is also independent of the control and detuning. 
 
\begin{figure}[tb]
\includegraphics[scale = 0.9]{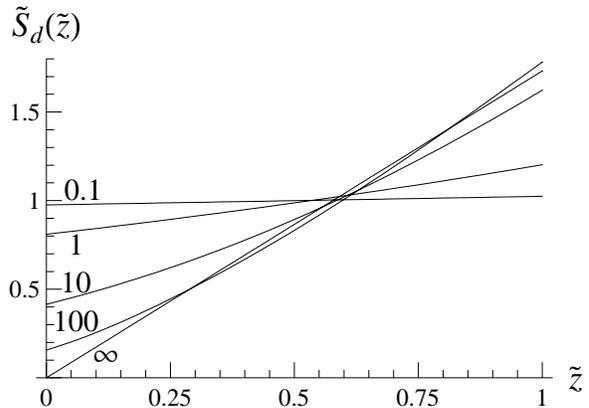}
\caption{Optimal modes $\tilde S_d(\tilde z)$ to retrieve from (in the forward direction) at indicated values of $d$. The flipped versions of these modes $\tilde S_d(1-\tilde z)$ are the optimal modes for backward retrieval and are also the optimal (normalized) spin waves $S(\tilde z,T)/\sqrt{\eta^\textrm{max}_\textrm{s}}$  for  adiabatic and fast storage if it is optimized for storage alone or for storage followed by backward retrieval ($\eta^\textrm{max}_\textrm{s}$ is the maximum storage efficiency). \label{fig:waves}}
\end{figure}

In contrast to the cavity case in paper I where there was only one spin-wave mode available, in the free-space case, the retrieval efficiency in Eq.~(\ref{etargen}) is different for different spin-wave modes. We can, thus, at each $d$, optimize retrieval by finding the optimal retrieval spin wave $\tilde S_d(\tilde z)$ (we suppress here the argument $\tilde T_\textrm{r}$). The expression for the efficiency in the last line of Eq.~(\ref{etargen}) is an expectation value of a real symmetric (and hence Hermitian) operator $k_\textrm{r}(\tilde z, \tilde z')$ in the state $S(1-\tilde z)$. It is therefore maximized when $S(1-\tilde z)$ is the eigenvector with the largest eigenvalue of the following eigenvalue problem:
\begin{equation} \label{eigeneqret}
\eta_\textrm{r} S(1-\tilde z) = \int_0^1 d \tilde z' k_\textrm{r}(\tilde z,\tilde z') S(1-\tilde z').
\end{equation} 
Since eigenvectors of real symmetric matrices can be chosen  real, the resulting optimal spin wave $\tilde S_d(\tilde z)$ can be chosen real. To find it, we start with a trial $S(\tilde z)$ and iterate the integral in Eq.~(\ref{eigeneqret}) several times until convergence \cite{mikhlin57}. In Fig.~\ref{fig:waves}, we plot the resulting optimal spin wave $\tilde S_d(\tilde z)$ for $d = 0.1, 1, 10, 100$, as well as its limiting shape ($\tilde S_\infty(\tilde z) = \sqrt{3} \tilde z$) as $d \rightarrow \infty$. At $d \rightarrow 0$, the optimal mode approaches a flat mode. These shapes can be understood by noting that retrieval is essentially an interference effect resembling superradiance, where the emission from all atoms contributes coherently in the forward direction. To get the maximum constructive interference, it is desirable that all atoms carry equal weight and phase in the spin wave. In particular, at low optical depth, this favors the flat spin wave. 
On the other hand, it is also desirable not to have a sudden change in the spin wave (except near the output end of the ensemble). The argument above essentially shows that excitations can decay through two different paths: by spontaneous emission in all directions or by collective emission into the forward direction. In Eq.~(\ref{freeeqs2p}), these two paths are represented by the $-P$ and $i\sqrt{d}\eop$ terms, respectively. The latter gives rise to a decay because Eq.~(\ref{freeeqs2e}) can be integrated to give a term proportional to $P$: $\eop= i \int d\tilde{z}\sqrt{d}P$. To obtain the largest decay in the forward direction all atoms should ideally be in phase so that the phase of $P(\tilde{z})$ is the same at all $\tilde{z}$. This constructive interference, however, is not homogeneous but builds up through the sample. At $\tilde{z}=0$, we have $\eop=0$, and the spontaneous emission is, thus, the only decay channel, i.e.,  $d|P(\tilde{z}=0)|^2/dt=-2|P(\tilde{z}=0)|^2$. To achieve the largest  retrieval efficiency, we should therefore put a limited amount of the excitation near $\tilde{z}=0$ and only have a slow build up of the spin wave from $\tilde{z}=0$ to $\tilde{z}=1$. The optimal spin-wave modes in Fig.~\ref{fig:waves} represent the optimal version of this slow build up. We will also reinterpret these optimal modes from a different perspective in Sec.~\ref{sec:freeadret} using the EIT window concept. 

From the qualitative argument given here, one can estimate the dependence of the optimal retrieval efficiency on the optical depth: we consider the emission into a forward mode of cross sectional area $A$. In the far field, this corresponds to a field occupying a solid angle of $\lambda^2/A$, where $\lambda = (2 \pi c)/\omega_1$ is the wavelength of the carrier. A single atom will, thus, decay into this mode with a probability $\sim \lambda^2/A$. With $N$ atoms contributing coherently in the forward direction, the emission rate is increased by a factor of $N$ to $\gamma_\textrm{f} \sim \gamma N \lambda^2/A$. The retrieval efficiency can then be found from 
the rate of desired ($\gamma_\textrm{f}$) and undesired ($\gamma$) decays as  $\eta=\gamma_\textrm{f}/(\gamma_\textrm{f}+\gamma)\sim 1- A/(N\lambda^2)$. By noting that $\lambda^2$ is the cross section for resonant absorption of a two-level atom, we recognize $N\lambda^2/A$ as the optical depth $d$ (up to a factor of order 1). The efficiency is then $\eta\sim 1- 1/d$, which is in qualitative agreement with the results of the full optimization which gives $1-\eta\approx 2.9/d$. A more detailed discussion of the dependence of retrieval efficiency on the shape of the spin wave is postponed until Sec.~\ref{sec:freeadret}, where Eq.~(\ref{etargen}) is rederived in the adiabatic limit and discussed in the context of the EIT window.

\section{Optimal Storage From Time Reversal: General Proof \label{sec:timerev}}

As already mentioned in paper I, the concept of time reversal allows us to deduce the optimal control strategy for storage from retrieval. In this section, we prove this result both for the free-space case and for the cavity case (the cavity case differs only in that there is just one spin-wave mode involved).  In the next section, we generalize these ideas and show that time reversal can be generally used to optimize state mappings. 

Despite the fact that our system contains nonreversible decay $\gamma$, time reversal is still an important and meaningful concept: to make time reversal applicable in this situation, we expand our system so that we not only consider the electric field and the spin wave, but also include all the reservoir modes, into which the excitations may decay. As discussed in Sec.~\ref{sec:freespacemodel} and Appendix \ref{sec:appModelfree}, the initial state of the reservoir modes is vacuum. When considering ``all the modes in the universe" \cite{scully97}, we have a closed system  described by (possibly infinitely many) bosonic creation operators $\{\hat O^\dagger_i\}$ with commutation relations
\begin{equation} \label{dcommrel}
\left[\hat O_i, \hat O^\dagger_j\right] = \delta_{i j}. 
\end{equation}
 
The evolution we consider here can be seen as a  generalized beam-splitter transformation, and can equivalently be specified as a Heisenberg picture map between the annihilation operators $\hat O_{i,\textrm{out}}= \sum_j U_{ij}[T,0;\Omega(t)] \hat O_{j,\textrm{in}}$ or as a Schr{\"o}dinger picture map $\hat U[T,0;\Omega(t)] = \sum_{i j} U_{ij}[T,0;\Omega(t)] |i\rangle \langle j|$ in the Hilbert space $\mathcal{H}$ with an orthonormal basis of single excitation states $|i\rangle=\hat O^\dagger_i |\textrm{vacuum}\rangle$. To stress that the mapping depends on the classical control field  $\Omega(t)$, we  here  include the argument $\Omega(t)$  in the evolution operator $\hat U[\tau_2,\tau_1;\Omega(t)]$, which takes the state from time $\tau_1$ to $\tau_2$. The operator $\hat U[T,0;\Omega(t)]$ must be unitary $\hat U^\dagger[T,0;\Omega(t)]=\hat U^{-1}[T,0;\Omega(t)]=\hat U[0,T;\Omega(t)]$. For simplicity of notation, we will here use the Schr{\"o}dinger picture.

Let us define two subspaces of $\mathcal{H}$: subspace $A$ of ``initial" states and subspace $B$ of ``final states." $B^\perp$, the orthogonal complement of $B$, can be thought of as the subspace of ``decay" modes (that is, the reservoir and other states, possibly including the ``initial" states, to which we do not want the initial states to be mapped). In this section, we will use $\hat U$ as the retrieval map, in which case $A$ and $B$ are spin-wave modes and output photon modes, respectively, while $B^\perp$ includes $A$, empty input field modes, and the reservoir modes, to which the excitations can decay by spontaneous emission. 

In the cavity derivation in paper I, we solved in the adiabatic limit for the control pulse shape $\Omega_{\textrm{r}}(t)$, which retrieves the atomic excitation into a specific mode $e(t)$. We then derived the pulse shape $\Omega_{\textrm{s}}(t)$, which optimally stores an incoming mode $\eop_{\textrm{in}}(t)$, and  noted that if the incoming mode is the time-reverse of the mode, onto which we retrieved, i.e.,  $\eop_{\textrm{in}}^* (T-t)=e(t)$, then the optimal storage control is the time-reverse of the retrieval control, i.e., $\Omega_{\textrm{s}}^*(T-t)=\Omega_{\textrm{r}}(t)$. Furthermore, in this case the storage and retrieval efficiencies were identical. As we now show, this is not a coincidence, but a very general result.

For the free-space case, we define the ``overlap efficiency" for storing into any given mode $S(z)$ as the expectation value of the number of excitations in this mode $S(z)$. Since the actual mode (call it $S'(z)$), onto which the excitation is stored, may contain components orthogonal to $S(z)$, the overlap efficiency for storing into $S(z)$ is in general less than the (actual) storage efficiency, and is equal to it only if $S(z) = S'(z)$.   

We will now prove that storing the time reverse of the output of backward retrieval from $S^*(z)$ with the time reverse of the retrieval control field gives the overlap storage efficiency into $S(z)$ equal to the retrieval efficiency. To begin the proof, we note that the probability to convert under $\hat U$ an initial excitation from a state $|a\rangle$ in A into a state  $|b\rangle$ in B is just
\begin{equation}
\label{etaUinv}
\eta=|\langle b |\hat U[T,0;\Omega(t)]|a\rangle|^2=|\langle a |\hat U^{-1}[T,0;\Omega(t)]|b\rangle|^2,
\end{equation}
where, in the last expression, we have used the unitarity of $\hat U[T,0;\Omega(t)]$. We now assume that $\hat U[T,0;\Omega(t)]$ describes retrieval and that $|a\rangle$ stands for $S^*(z)$, while $|b\rangle$ stands for the output field mode $\eop$, onto which $S^*$ is retrieved under $\Omega(t)$. Then $\eta$ is just the retrieval efficiency from $S^*$. The last expression then shows that if we could physically realize the operation $\hat U^{-1}[T,0;\Omega(t)]$, then it would give the overlap efficiency for storage of $\eop$ into $S^*$ equal to the retrieval efficiency $\eta$. The challenge is, therefore, to physically invert the evolution and realize $\hat U^{-1}[T,0;\Omega(t)]$. As we now show, time reversal symmetry allows us to perform this inverse evolution in some cases. We refer the reader to Appendix \ref{sec:apptimerev} for a careful definition of the time reversal operator $\hat{\mathcal{T}}$ and for the proof of the following equality:
\begin{equation} \label{inverseprop}
\hat U^{-1}[T,0;\Omega(t)]  = \hat{\mathcal{T}} \hat U[T,0;\Omega^*(T-t)] \hat{\mathcal{T}}, 
\end{equation}
where it is implicit that the carrier wave vector of the time-reversed control pulse $\Omega^*(T-t)$ propagates in the direction opposite to the carrier of $\Omega(t)$.  Physically, Eq.~(\ref{inverseprop}) means that we can realize the inverse evolution 
by time-reversing the initial state, evolving it using a time-reversed control pulse, and finally time-reversing the final state. Then, using Eqs.~(\ref{etaUinv}) and (\ref{inverseprop}), the retrieval efficiency may be rewritten as
\begin{equation} \label{etarev1}
\eta = |\langle a |\hat{\mathcal{T}} \hat U[T,0;\Omega^*(T-t)] \hat{\mathcal{T}}|b\rangle|^2.
\end{equation}
This means that if we can retrieve the spin wave $S^*$ backwards onto $\eop(t)$ using $\Omega(t)$, we can use $\Omega^*(T-t)$ to store $\eop_\textrm{in}(t)=\eop^*(T-t)$ onto $S$ with the overlap storage efficiency equal to the retrieval efficiency $\eta_\textrm{r}$. 

We will now prove that this time-reversed storage is also the optimal solution, i.e., that an overlap efficiency for storage into $S$ greater than $\eta_\textrm{r}$ is not possible. To begin the proof, let us suppose, on the contrary, that we can store $\eop_\textrm{in}(t)$ into $S$ with an overlap efficiency $\eta_\textrm{s} > \eta_\textrm{r}$. Applying now the time reversal argument to storage, we find that backward retrieval from $S^*$ with the time-reversed storage control will have efficiency greater than $\eta_\textrm{r}$. However, from Eq.~(\ref{etargen}), we know that the retrieval efficiency is independent of the control field and is invariant under the complex conjugation of the spin wave, so we have reached a contradiction. Therefore, the maximum overlap efficiency for storage into a given mode $S$ is equal to the backward retrieval efficiency from $S^*$ (and $S$) and can be achieved by time-reversing backward retrieval from $S^*$. 

Finally, the strategy for storing $\eop_\textrm{in}(t)$ with the maximum storage efficiency (rather than maximum overlap efficiency into a given mode, as in the previous
paragraph) follows immediately from the arguments above: provided we can retrieve the (real) optimal backward-retrieval mode $\tilde S_d(L-z)$  backwards into
$\eop_\textrm{in}^*(T-t)$, the optimal storage of $\eop_\textrm{in}(t)$ will be the time reverse
of this retrieval and will have the same efficiency
as the optimal retrieval efficiency at this $d$, i.e., the retrieval
efficiency from $\tilde S_d$.  

While the above argument is very general, it is important to realize its key limitation. The argument shows that it is possible to optimally store a field $\eop_\textrm{in}(t)$ provided we can optimally retrieve onto $\eop_\textrm{in}^*(T-t)$ (i.e., backward-retrieve $\tilde S_d(L-z)$ into $\eop_\textrm{in}^*(T-t)$). It may, however, not be possible to optimally retrieve onto $\eop_\textrm{in}^*(T-t)$ because it may, for example, be varying too fast. For this reason, we shall explore in Secs.~\ref{sec:freeadret} and \ref{sec:freefast}, onto which fields it is possible to retrieve a given spin wave. Before we do this, however, we will show in the next section that time reversal does not only allow one to derive the optimal storage strategy from the optimal retrieval strategy, as we did in this section, but also allows one to find the optimal spin wave for retrieval.   

\section{Time Reversal as a Tool for Optimizing Quantum State Mappings\label{sec:timerev2}}

We will now show that time reversal can be used as a general tool for optimizing state mappings. Moreover, we will show that for the photon storage problem considered in this paper, in addition to being a very convenient mathematical tool, the optimization procedure based on time reversal may also be realized experimentally in a straightforward way.  

In Sec.~\ref{sec:freeret} we found $\tilde S_d(z)$, the optimal spin wave to retrieve from, by starting with a trial spin wave $S_1(z)$ and iterating  Eq.~(\ref{eigeneqret}) until convergence. While we just used this as a mathematical tool for solving an equation, the iteration procedure actually has a physical interpretation. Suppose that we choose a certain classical control $\Omega(t)$ and retrieve the spin wave $S(z)$ forward onto $\eop(t)$ and then time reverse the control to store $\eop^*(T-t)$ backwards. By the argument in the last section, this will, in general, store into a different mode $S'(z)$ with a higher efficiency (since the actual storage efficiency is, in general, greater than the overlap storage efficiency into a given mode). In this way, we can iterate this procedure to compute spin waves with higher and higher forward retrieval efficiencies \cite{fox61, gerchberg72}. In fact,
forward retrieval followed by time-reversed backward storage can be expressed as 
\begin{equation} \label{generalizedeigeneqret}
S_2(1-\tilde z) = \int_0^1 d \tilde z' k_\textrm{r}(\tilde z,\tilde z') S^*_1(1-\tilde z'),
\end{equation}
which for real $S$ is identical to the iteration of Eq.~(\ref{eigeneqret}).
We note that the reason why backward storage had to be brought up here (in contrast to the rest of the paper, where storage is always considered in the forward direction) is because Eq.~(\ref{eigeneqret}), which Eq.~(\ref{generalizedeigeneqret}) is equivalent to for real $S$, discusses forward retrieval, whose time-reverse is backward storage. 

Since the iterations used to maximize the efficiency in Eq.~(\ref{eigeneqret}) are identical to Eq.~(\ref{generalizedeigeneqret}), the physical interpretation of the iterations in Eq.~(\ref{eigeneqret}) is that we retrieve the spin wave and store its time-reverse with the time-reversed control field (i.e., implement the inverse $\hat U^{-1}$ of the retrieval map $\hat U$ using Eq.~(\ref{inverseprop})). We will explain below that this procedure of retrieval followed by time-reversed storage can be described mathematically by the operator $\hat{\mathcal{N}} \hat P_A \hat U^{-1} \hat P_B \hat U$, where $\hat P_A$ and $\hat P_B$ are the projection operators on the subspaces $A$ and $B$ of spin wave modes and output photon modes, respectively, and where $\hat{\mathcal{N}}$ provides renormalization to a unit vector. It is, in fact, generally true that in order to find the unit vector $|a\rangle$ in a given subspace $A$ of ``initial" states that maximizes the efficiency $\eta= |\hat P_B \hat U |a\rangle|^2$ of a given unitary map $\hat U$ (where $B$ is a given subspace of ``final" states), one can start with any unit vector $|a\rangle \in A$ and repeatedly apply $\hat{\mathcal{N}} \hat P_A \hat U^{-1} \hat P_B \hat U$ to it. We prove in Appendix \ref{sec:apptimerev2} that this procedure converges to the desired optimal input mode $|a_\textrm{max}\rangle$ yielding the maximum efficiency $\eta_\textrm{max}$ (provided $\langle a|a_\textrm{max}\rangle \neq 0$) and that $\hat{\mathcal{N}} \hat P_B \hat U |a_\textrm{max}\rangle$ also optimizes $\hat U^{-1}$ as a map from $B$ to $A$.
 
We have just discussed the iterative optimization procedure as a purely mathematical tool for computing the optimal initial mode in the subspace $A$ of initial states. 
However, in the previous section and in Appendix \ref{sec:apptimerev}, we showed that for our system one can implement the inverse evolution $U^{-1}$ experimentally by first time reversing the state of the system, then applying ordinary evolution but with a time-reversed control, and then time reversing the state of the system again. Thus, in addition to being a convenient mathematical tool, the time reversal based optimization technique suggested in this paper is an experimentally realizable procedure. In fact, following this work, this procedure has recently been applied experimentally to the optimization of the storage and retrieval of light \cite{novikova07}.

As an example, let us discuss how this experimental implementation applies to the optimization of backward retrieval. The full Hilbert space is spanned by subspace $A$ of spin-wave modes, subspace $B$ of output field modes, as well as a subspace containing (empty) input and reservoir field modes. The goal is to optimize the retrieval time evolution map $\hat U[\Omega(t)]$ for some fixed detuning $\Delta$ and fixed backward-propagating Rabi frequency pulse $\Omega(t)$ (sufficiently powerful for complete retrieval) with respect to the initial spin wave $|a\rangle \in A$. From Sec.~\ref{sec:timerev}, it follows that the iteration $\hat{\mathcal{N}} \hat P_A \hat U^{-1}[\Omega(t)] \hat P_B \hat U[\Omega(t)] |a\rangle$ required for the optimization can be experimentally implemented as follows. We start with a spin-wave mode $|a\rangle$  with a real mode shape
$S(z)$, carry out backward retrieval, and measure the outgoing field.
We then prepare the time reverse of the measured field shape and
store it back into the ensemble using the time-reversed control pulse. The projections $\hat P_B$ and $\hat P_A$ happen automatically since we do not reverse the reservoir
modes and the leak. The renormalization can be achieved during the generation
of the time-reversed field mode, while the time reversal for the spin wave will be unnecessary since a real spin wave will stay real under retrieval followed by time-reversed storage. The iteration suggested here is, thus, indeed equivalent to the iteration in Eq.~(\ref{generalizedeigeneqret}) with $S(1-\tilde z)$ replaced with $S(\tilde z)$ (since Eq.~(\ref{generalizedeigeneqret}) optimizes forward retrieval). 

For single photon states, the measurement of the outgoing field involved in the procedure above
will require many runs of the experiment. To circumvent this, one can use the fact that the equations of motion (\ref{freeeqs2e})-(\ref{freeeqs2s}) for the envelope of the quantum field mode are identical to the equations of motion for the classical field propagating under the same conditions. One can, thus, use the optimization procedure with classical light pulses and find optimal pairs of input pulse shapes and control fields, which will give optimal storage into the spin wave $\tilde S_d(1-\tilde z)$. However, since the equations of motion for quantum light modes are identical to the classical propagation equations, this data can then be interpreted as optimal pairs of control fields and quantized input photon modes for optimal storage of nonclassical light (such as single photons) into the optimal backward retrieval mode $\tilde S_d(1-\tilde z)$. 
      
We will now briefly discuss the application of time reversal ideas to the optimization of the combined process of storage followed by retrieval. For real spin waves, storage and backward retrieval are time reverses of each other since real spin waves are unaltered by complex conjugation. Consequently, the time reversal iteration of storage and backward retrieval optimizes both of them, as well as the combined process of storage followed by backward retrieval. Therefore, for a given input, the storage control field that optimizes storage alone will also be optimal for storage followed by backward retrieval.

In contrast, (forward) storage and forward retrieval are not time reverses of each other, and the entire process of storage followed by forward retrieval has to be optimized as a whole. The general time reversal iteration procedure can still be used in this case with the understanding that the spaces $A$ and $B$ of initial and final states are the right propagating modes to the left of the ensemble (except for later empty input modes during retrieval) and right propagating modes to the right of the ensemble (except for earlier storage leak modes), respectively, while the remaining modes are reservoir modes, spin-wave modes, leak modes, and empty incoming photon modes from the left during retrieval. Since the time-reverse of storage followed by forward retrieval is itself storage followed by forward retrieval except in the opposite direction, the optimization can be carried out physically by starting with a given input field mode, storing it and retrieving it forward with given control pulses, time reversing the output and the control pulses, and iterating the procedure. The optimal control-dependent input field, which the iteration will converge to, will then be stored into a particular optimal spin wave, which itself will be independent of the control used for the iteration. In Secs.~\ref{sec:freeadstret} and \ref{sec:freefast} we will look at storage followed by forward retrieval in more detail.   

It is important to note that the discussion in this section assumed that the two metastable states are degenerate. If they are not degenerate, a momentum $\Delta k = \omega_{sg}/c$ will be written onto the spin wave during storage, so that its time reversal will no longer be trivial. In Sec.~\ref{sec:nondeg}, we will discuss in detail how the optimization is modified when the metastable states are not degenerate.

Procedures that are related to ours and that also use time-reversal iterations 
for optimization are a standard tool in applied optimal control \cite{krotov96, bryson75, krotov83, konnov99} and have been used for a variety of applications in chemistry \cite{shapiro03, kosloff89}, NMR \cite{khaneja05}, and atomic physics \cite{sklarz02, calarco04}. In most of these works, time reversal iterations are used as a mathematical tool for computing, for a given initial state $|a\rangle$, the optimal time-dependent control that would result in the final state with the largest projection on the desired subspace $B$ of final states. In fact, this mathematical tool is directly applicable to our problem of shaping the control pulses, as we will discuss elsewhere \cite{gorshkov07b}. However, our use of time reversal iterations in the present paper and in papers I and III differs in two ways from that of Refs.\ \cite{krotov96, bryson75, krotov83, konnov99,shapiro03, kosloff89,khaneja05,sklarz02, calarco04}. First, we use time reversal iterations to find the optimal $|a\rangle$ in the subspace $A$ of initial states for a given propagator $U[\Omega(t)]$, rather than to shape the control $\Omega(t)$ itself for a given $|a\rangle$ (we shape the control by explicitly solving the equations, as explained in Secs.~\ref{sec:freead} and \ref{sec:freefast}). Second, the time reversal iterations discussed in the references above are a purely mathematical technique, while our iterative algorithm can be implemented experimentally \cite{novikova07}.

The main result of this section is an iterative procedure for solving or experimentally finding the optimal retrieval spin wave, while the main result of the previous section was that one can generate optimal pairs of inputs and control fields by time-reversing the output and the control field of such optimal retrieval. In order to say, however, for which input fields the optimal storage control $\Omega(t)$ can be found (or, equivalently, into which output fields can the optimal spin-wave excitation be retrieved), we need to consider the limits, in which Eq.~(\ref{freeoneeq}) can be fully solved analytically. These limits, adiabatic and fast, will be discussed in Secs.~\ref{sec:freead} and \ref{sec:freefast}, respectively.    
  
\section{Adiabatic Retrieval and Storage \label{sec:freead}}
   
\subsection{Adiabatic retrieval \label{sec:freeadret}}

\subsubsection{Solution for the output field and for the retrieval efficiency\label{sec:freeadretsol}}

Based on the branching ratio and the time reversal arguments, we have found the maximal storage efficiency at each $d$ and have described the optimal storage scenario in the three preceding sections. Since a given input mode can be optimally stored if and only if optimal retrieval can be directed into the time-reverse of this mode, in the following sections (Secs.~\ref{sec:freead} and \ref{sec:freefast}), we solve Eq.~(\ref{freeoneeq}) analytically in two important limits to find out, which modes we can retrieve into and store optimally. The first of these two limits, which we will consider in the next five sections (Secs.~\ref{sec:freeadret} - \ref{sec:freespindecay}), corresponds to smooth input and control  fields, such that the term in the square brackets in Eq.~(\ref{freeoneeq}) can be ignored. This ``adiabatic limit'' corresponds to an adiabatic elimination of the optical polarization $P$ in Eq.~(\ref{freeeqs2pl}). The precise conditions for this adiabatic elimination will be discussed in Sec.~\ref{sec:freeadcond}. In this section (Sec.~\ref{sec:freeadret}), we consider the retrieval process.

Similar to the cavity discussion in Sec.\ V A of paper I, it is instructive to note that in the adiabatic approximation (i.e., with $\partial_{\tilde t} P$ in Eq.~(\ref{freeeqs2p}) replaced with $0$), rescaling variables $\eop$ and $P$ by $\tilde \Omega$ and changing variables $\tilde t \rightarrow h(\tilde T_\textrm{r}, \tilde t)$, where (as in Eq.~(15) in paper I, except now in dimensionless form)
\begin{equation} \label{hdeffree}
h(\tilde t,\tilde t') = \int_{\tilde t}^{\tilde t'} |\tilde \Omega(\tilde t'')|^2 d \tilde t'',
\end{equation}  
makes Eqs.~(\ref{freeeqs2e})-(\ref{freeeqs2s}) independent of $\Omega$. This allows one to solve these equations in an $\Omega$-independent form and then obtain the solution for any given $\Omega$ by simple rescaling. A special case of this observation has also been made in Ref.~\cite{nunn06}, where the authors treat the Raman limit. However, since Eqs.~(\ref{freeeqs2e})-(\ref{freeeqs2s}) are relatively simple, we will avoid causing confusion by using new notation and will solve these equations without eliminating $\Omega$.

To solve for the output field during adiabatic retrieval, we assume for simplicity that retrieval begins at time $\tilde t = 0$ rather than at time $\tilde t = \tilde T_\textrm{r}$ and that the initial spin wave is $S(\tilde z,\tilde t = 0) = S(\tilde z)$. In the adiabatic approximation, Eqs.~(\ref{freeeqs2s}) and (\ref{freeeqs2pl}) reduce to a linear first order ordinary differential equation on $S$.  Solving this equation for $S(u,\tilde t)$ in terms of $S(u')$, expressing $\eop(u,\tilde t)$ in terms of $S(u, \tilde t)$ using Eqs.~(\ref{freeeqs2el}) and (\ref{freeeqs2pl}), and taking the inverse Laplace transform $u \rightarrow \tilde z = 1$, we arrive at 
\begin{eqnarray} \label{freeadeout}
\eop(1,\tilde t) &=& - \sqrt{d} \tilde \Omega(\tilde t) \int_0^1 d \tilde z \frac{1}{1 + i \tilde \Delta} e^{-\frac{h(0,\tilde t) + d \tilde z}{1 + i \tilde \Delta}} 
\nonumber \\
&& \times  I_0\left(2 \frac{\sqrt{h(0,\tilde t) d \tilde z}}{1+i \tilde \Delta}\right) S(1-\tilde z). 
\end{eqnarray}
The $\tilde t$-dependent and the $\tilde z$-dependent phases in the exponent can be interpreted as the ac Stark shift and the change in the index of refraction, respectively.

By using the identity \cite{gray1895}
\begin{equation} \label{besselequality}
\int_0^\infty dr r e^{-p r^2} I_0(\lambda r) I_0(\mu r) = \frac{1}{2 p} e^{\frac{\lambda^2 + \mu^2}{4 p}} I_0\left(\frac{\lambda \mu}{2 p}\right) 
\end{equation}
for appropriate $\mu$, $\lambda$, and $p$, we find that 
for a sufficiently large $h(0,\infty)$ ($d h(0,\infty) \gg |d + i \tilde \Delta|^2$), 
 the retrieval efficiency (Eq.~(\ref{etar}) with $\tilde T_\textrm{r}$ replaced with $0$) is
\begin{equation} \label{freereteff}
\eta_\textrm{r} = \int_0^1 d \tilde z \int_0^1 d \tilde z' k_\textrm{r}(\tilde z, \tilde z') S(1-\tilde z) S^*(1-\tilde z'),
\end{equation}
in agreement with Eq.~(\ref{etargen}). So $\eta_\textrm{r}$ is independent of detuning and control pulse shape but depends on the spin wave and the optical depth. Thus, the adiabatic approximation does not change the exact value of efficiency and keeps it independent of  detuning and classical control field by preserving the branching ratio between desired and undesired state transfers. It is also worth noting that, in contrast to Eq.~(\ref{etargen}), Eq.~(\ref{freeadeout}) allows us to treat and optimize retrieval even when the energy in the control pulse is limited \cite{nunn06} (i.e., $d h(0,\infty) \lesssim |d + i \tilde \Delta|^2$). However, in the present paper, the treatment of adiabatic retrieval is focused on the case when the control pulse energy is sufficiently large ($d h(0,\infty) \gg |d + i \tilde \Delta|^2$) to leave no excitations in the atoms and to ensure the validity of Eq.~(\ref{freereteff}) (or, equivalently, Eq.~(\ref{etargen})). 

As pointed out in the Introduction to paper I, two important subsets of the adiabatic limit, the resonant limit and the Raman limit, are often considered in the literature because the equations can be simplified in these limits. Although we demonstrate in this work that the basic underlying physics and hence the optimal performance are the same for these two photon storage techniques, a more detailed analysis reveals significant differences. It is precisely these differences that obstruct the underlying equivalence between the two protocols. And it is these differences that make this equivalence remarkable. As an example of such a difference, resonant and Raman limits give different dependence on $d$ of the duration $T_\textrm{out}$ of the output pulse. To see this, it is convenient to ignore the decay in Eq.~(\ref{freeadeout}) (due to the rescaling, this means we ignore $1$ in $1+i\tilde \Delta$). If we do this, we obtain
\begin{eqnarray} \label{freeadeoutdecayless}
\eop(1,\tilde t) &=& i \sqrt{d} \tilde \Omega(\tilde t) \int_0^1 d z \frac{1}{\tilde \Delta} e^{i \frac{h(0,\tilde t) + d \tilde z}{\tilde \Delta}} J_0\left( 2\frac{\sqrt{h(0,\tilde t) d \tilde z}}{\tilde \Delta}\right) 
\nonumber \\
&&\times S(1-\tilde z), 
\end{eqnarray}
where $J_0(x) = I_0(i x)$ is the zeroth order Bessel function of the first kind.
In the resonant limit ($d \gamma \gg |\Delta|$), we find the duration of the output pulse by observing that the $\tilde \Delta \rightarrow 0$ limit of Eq.~(\ref{freeadeoutdecayless}) is
\begin{equation} \label{gpvelprop}
\eop(1,\tilde t) = -\frac{\tilde \Omega(\tilde t)}{\sqrt{d}} S\left(1- \frac{h(0,\tilde t)}{d}\right),
\end{equation}   
with the understanding that $S(\tilde z)$ vanishes outside of $[0,1]$. This is just the ideal lossless and dispersionless group velocity propagation of Refs.~\cite{fleischhauer02,lukin00fleischhauer00}, also know as EIT. This implies a duration $T_\textrm{out} \sim d \gamma/|\Omega|^2$ for the output pulse in the resonant limit, which is consistent with the cavity case analyzed in paper I if one identifies $C$ and $d$. In the Raman limit ($d \gamma \ll |\Delta|$), the length of the output pulse is simply given by the fall-off of $J_0$ and is found from $h(0,\tilde  t) d/\tilde \Delta^2 \sim 1$ to be $T_\textrm{out} \sim \Delta^2/(\gamma d |\Omega|^2)$. It is worth noting that the appropriate Raman limit condition is $\gamma d \ll |\Delta|$ and not $\gamma \ll |\Delta|$ as one might naively assume by analogy with the single-atom case. It is also important to note that if one is limited by laser power (as in Ref.~\cite{nunn06}) and desires to achieve the smallest possible $T_\textrm{out}$, the above formulas for $T_\textrm{out}$ imply that EIT retrieval is preferable over Raman. 

\subsubsection{Dependence of retrieval error on optical depth $d$\label{sec:retddep}}

In the cavity case analyzed in paper I, only one spin-wave mode is available and the retrieval error is always $1/(1+C)$ ($\approx 1/C$ for $C \gg 1$). In free space, in contrast, the retrieval error depends on the spin-wave mode $S(\tilde z)$ and as we will explain in this section, scales differently with $d$ depending on the spin wave. Since the retrieval efficiency is independent of $\Delta$, to gain some physical intuition for the error dependence on the spin wave and on $d$, we will focus on the $\Delta = 0$ case, for which the formalism of EIT transparency window has been developed \cite{fleischhauer05}. For $\Delta = 0$ and large $d$, the integrand in Eq.~(\ref{freeadeout}) can be approximated with a Gaussian. Then using dimensionless momentum $\tilde k = k L$ and defining the Fourier transform of $S(\tilde z)$ as $S(\tilde k) = (2 \pi)^{-1} \int_0^1 d \tilde z S(\tilde  z) \exp(-i \tilde k \tilde  z)$, we can write Eq.~(\ref{freeadeout}) as
\begin{equation} \label{gaussian1}
\eop(1,\tilde  t) = - \frac{\tilde  \Omega(\tilde  t)}{\sqrt{d}}\int_{-\infty}^{\infty} d \tilde k   e^{i \tilde k \left(1-\frac{h(0,\tilde  t)}{d}\right)} e^{- h(0,\tilde  t) \frac{\tilde k^2}{d^2}} S(\tilde  k).  
\end{equation}
In the limit $d \rightarrow \infty$, the Gaussian term can be replaced with $1$ to yield back the group velocity propagation in Eq.~(\ref{gpvelprop}). Computing the efficiency using Eq.~(\ref{gaussian1}), we find, after a change of variables $\tilde t \rightarrow \tau = h(0,\tilde t)/d$,
\begin{equation} \label{gaussianeffic}
\eta_\textrm{r} = \int_0^\infty d \tau \left| \int_{-\infty}^{\infty} d \tilde  k    e^{i \tilde k (1 - \tau)} e^{- \frac{\tilde k^2}{d/\tau}} S(\tilde k)\right|^2.  
\end{equation}

We will now show that the Gaussian term of width $\Delta \tilde k_\textrm{EIT} = \sqrt{d/\tau}$ in the integrand in Eq.~(\ref{gaussianeffic}) can be interpreted as the effective momentum-space EIT transparency window for the spin wave. We start by noting that the equivalent of $\tau$ in the original units (call it $z_\textrm{prop}$) is equal to $z_\textrm{prop} = L \tau(t) = \int_0^t v_\textrm{g}(t') d t'$ and, thus, represents the propagation distance ($v_\textrm{g} = c \Omega^2/(g^2 N)$ is the EIT group velocity \cite{fleischhauer05}). This interpretation of $z_\textrm{prop}$ also follows from the fact that, if one ignores the Gaussian in Eq.~(\ref{gaussianeffic}), the spin wave would be evaluated at $\tilde z = 1 - \tau$. Thus, in terms of the propagation distance $z_\textrm{prop}$, the width of the momentum-space transparency window in Eq.~(\ref{gaussianeffic}) can be written, in original units, as $\Delta k_\textrm{EIT} = \Delta \tilde k_\textrm{EIT}/L = \sqrt{d/\tau}/L = \sqrt{g^2 N/(\gamma c z_\textrm{prop})}$. Thus, as the propagation distance $z_\textrm{prop}$ decreases, the width $\Delta k_\textrm{EIT}$ of the transparency window gets wider and eventually becomes infinite at the $\tilde z = 1$ end of the ensemble, where $z_\textrm{prop} = 0$. The consistency of our expression $\Delta k_\textrm{EIT}$ for the effective momentum-space EIT window with the expression for the frequency-space EIT transparency window  $\Delta \omega_\textrm{EIT} = v_\textrm{g} \sqrt{g^2 N/(\gamma c z_\textrm{prop})}$ \cite{fleischhauer05} immediately follows from rescaling by $v_\textrm{g}$ both $\Delta \omega_\textrm{EIT}$ and the dark state polariton \cite{fleischhauer02,lukin00fleischhauer00} bandwidth $\Delta \omega_\textrm{p} = v_\textrm{g} \Delta k_\textrm{spin}$ (where $\Delta k_\textrm{spin}$ is the width of $S(k)$ in the original units). In fact, the change of variables $t \rightarrow \tau$ that led to Eq.~(\ref{gaussianeffic}) precisely accomplished this rescaling of the polariton and the EIT window by the group velocity. It is worth noting that this proportionality of both the polariton bandwidth and the frequency-space EIT window width to the group velocity (and, hence, the existence of the control-independent effective momentum-space EIT window) is another physical argument for the independence of retrieval efficiency from the control power.

An important characterization of the performance of an ensemble-based memory is the scaling of error with optical depth at large optical depth. In the cavity case analyzed in paper I, there was only one spin-wave mode available and the retrieval error for it was $1/(1+C)$ ($\approx 1/C$ for $C \gg 1$), where the cooperativity parameter $C$ can be thought of as the effective optical depth enhanced by the cavity. By qualitative arguments, we showed in Sec.~\ref{sec:freeret} that the retrieval efficiency in free space is $1-\eta \sim 1/d$. A more precise value can be found numerically from the optimal spin wave (Sec.~\ref{sec:freeret}), which gives a maximal retrieval efficiency that scales approximately as $\sim 2.9/d$, i.e., one over the first power of density, precisely as in the cavity case. However, this $1/d$ scaling turns out to be not the only possibility in free space. The scaling of the retrieval error with $d$ can be either $1/d$ or $1/\sqrt{d}$ depending on the presence of steps (i.e., discontinuities in the amplitude or phase of the spin wave). Specifically, numerics show that for a spin wave that does not have steps at any $\tilde z < 1$, the retrieval error scales as $1/d$, while steps in the phase or amplitude of the spin wave result in a $1/\sqrt{d}$ error. In particular, a step in the amplitude of $S(\tilde z)$ at position $\tilde z$ of height $l$ can be found numerically to contribute an error of $l^2 \sqrt{2/\pi} \sqrt{1-\tilde z}/\sqrt{d}$ at large $d$. The reason for the importance of steps is that at high $d$, the effective EIT window is very wide and only the tails of the Fourier transform $S(\tilde k)$ of the spin wave $S(\tilde z)$ matter. A step, i.e., a discontinuity, in the function $S(\tilde z)$ means that its Fourier transform falls off as $S(\tilde k) \sim 1/\tilde k$. Thus, if we assume all frequencies outside of an effective EIT window of width $\Delta \tilde k_\textrm{EIT}=\sqrt{d/\tau} \sim \sqrt{d/(1-\tilde z)}$ get absorbed, the error will be proportional to $\int^\infty_{\Delta \tilde k_\textrm{EIT}} d \tilde k |S(\tilde k)|^2 \sim \sqrt{(1-\tilde z)/d}$, precisely as found with numerics. Numerics also show that if a step in $|S(\tilde z)|$ is not infinitely sharp, at a given $d$, a feature should be regarded as a step if the slope of $|S(\tilde z)|^2$ is bigger than $\sim \sqrt{d}$ (and will then contribute a $1/\sqrt{d}$ error). The simple physical reason for this is that only if a feature in $S(\tilde z)$ is narrower than $1/\sqrt{d}$ will it extend in $\tilde k$ space outside the effective EIT window of width $\Delta \tilde k_\textrm{EIT} = \sqrt{d/\tau}$. While we only performed detailed analysis of steps in the amplitude of $S(\tilde z)$, steps in the phase of $S(\tilde z)$, as we have already noted, also contribute a $1/\sqrt{d}$ error, and we expect that similar dependence on the position and sharpness of such phase steps holds.   

A useful analytical result on scaling that supports these numerical calculations is the error on retrieval from a flat spin wave $S(\tilde z) = 1$, which can be calculated exactly from Eq.~(\ref{freereteff})  to be 
\begin{equation}
1-\eta_\textrm{r} = e^{-d} (I_0(d) + I_1(d)).
\end{equation}
Using the properties of modified Bessel functions of the first kind, one finds that as $d \rightarrow \infty$, the error approaches $\sqrt{2/\pi}/\sqrt{d}$, which is consistent with the general formula since a flat spin wave has one step at $\tilde z < 1$, i.e., a step of height $1$ at $\tilde z = 0$. In fact, it is this analytical result that allows us to exactly identify the $\sqrt{2/\pi}$ prefactor in the error due to amplitude steps.

Based on the results of this section and the results of paper I, we can identify several advantages of using a cavity setup. First, in the cavity, the optical depth is enhanced by the value of the cavity finesse from the free-space value of $d$ to form the cooperativity parameter $C$. Moreover, in terms of $d$ and $C$ the errors during optimal storage in free space and in the cavity scale as $2.9/d$ and $1/C$, respectively. That is, even if one ignores the enhancement due to cavity finesse, the cavity offers a factor of $3$ improvement. In addition to that, if one is forced to retrieve from the flat spin-wave mode $S(\tilde z)=1$ (which is the case, for example, if the spin wave is generated via spontaneous Raman scattering as in Ref.~\cite{eisaman05}), the free-space error is increased from the optimal and scales as $\sqrt{2/\pi}/\sqrt{d}$, while in the cavity case the mode $S(\tilde z)=1$ is, in fact, the only mode coupled to the cavity mode and is, therefore, precisely the one that gives $1/C$ scaling. Finally, because there is only one spin wave mode accessible in the cavity setup, the time reversal based iterative optimization procedure (Sec.~\ref{sec:timerev2}) requires only one iteration in the cavity case. On the other hand, the free-space setup described in this paper is much simpler to realize in practice and allows for the storage of multiple pulses in the same ensemble, e.g., time-bin encoded qubits \cite{afzelius06b}.  

Using the effective EIT window concept developed in this section, we can now interpret from a different perspective the optimal retrieval spin waves computed in Sec.~\ref{sec:freeret}. These spin waves represent at each $d$ the optimal balance between maximal smoothness (to minimize the momentum space width $\Delta \tilde k_\textrm{spin}$ of the spin wave so that it better fits inside the effective EIT window) and least amount of propagation (to minimize $\tau$ and, thus, maximize the width $\Delta \tilde k_\textrm{EIT} = \sqrt{d/\tau}$ of the effective EIT window itself).

\subsubsection{Shaping retrieval into an arbitrary mode\label{sec:retshaping}}

We have shown that optimal storage of a given input mode requires the ability to retrieve optimally into the time reverse of this input mode. Thus, by finding the modes we can  retrieve into, we will also find  the modes that  can be optimally stored. In this section we prove that by adjusting the control during retrieval  we can retrieve from any mode $S(\tilde z)$ into any given normalized mode $\eop_2(\tilde t)$, provided the mode is sufficiently smooth to satisfy the adiabaticity condition (which in the original units means $T_\textrm{out} d \gamma \gg 1$, where $T_\textrm{out}$ is the duration of $\eop_2(t)$, as we discuss in Sec.~\ref{sec:freeadcond}) \cite{patnaik04}. 

We know from Sec.~\ref{sec:freeret} that the retrieval efficiency $\eta_\textrm{r}$ is independent of the detuning $\Delta$ and the control $\Omega$, provided the retrieval is complete (for adiabatic retrieval, the condition on the control pulse energy for complete retrieval is $d h(0,\infty) \gg |d + i \tilde \Delta|^2$, as found in Sec.~\ref{sec:freeadretsol}). Thus, to find the control that retrieves $S(\tilde z)$ into any given normalized mode $\eop_2(\tilde t)$ with detuning $\tilde \Delta$, we need to solve for $\tilde \Omega(\tilde t)$ in equation (\ref{freeadeout}) with $\eop(1,\tilde t) = \sqrt{\eta_\textrm{r}} \eop_2(\tilde t)$:
\begin{eqnarray} \label{etae2} 
\sqrt{\eta_\textrm{r}} \eop_2(\tilde t) &=& - \sqrt{d} \tilde \Omega(\tilde t) \int_0^1 d \tilde z \frac{1}{1 + i \tilde \Delta} e^{-\frac{h(0,\tilde t) + d \tilde z}{1 + i \tilde \Delta}} 
\nonumber \\
&& \times  I_0\left(2 \frac{\sqrt{h(0,\tilde t) d \tilde z}}{1+i \tilde \Delta}\right) S(1-\tilde z). 
\end{eqnarray}
To solve for $\tilde \Omega(\tilde t)$, we integrate the norm squared of both sides from $0$ to $\tilde t$ and change the integration variable from $\tilde t'$ to $h'=h(0,\tilde t')$ on the right hand side to obtain
\begin{eqnarray}\label{hequation}
\eta_\textrm{r} \int_0^{\tilde t} d \tilde t' |\eop_2(\tilde t')|^2 & = & \int_0^{h(0,\tilde t)} d h' \Bigg|\int_0^1 d \tilde z \frac{\sqrt{d}}{1 + i \tilde \Delta} e^{-\frac{h' + d \tilde z}{1 + i \tilde \Delta}} 
\nonumber \\
&& \times  I_0\left(2 \frac{\sqrt{h' d \tilde z}}{1+i \tilde \Delta}\right) S(1-\tilde z)\Bigg|^2. 
\end{eqnarray}
In the cavity case, the corresponding equation (20) of paper I was solvable analytically. To solve Eq.~(\ref{hequation}) for $h(0,\tilde t)$ numerically, we note that both sides of Eq.~(\ref{hequation}) are monotonically increasing functions of $\tilde{t}$ and are equal at $\tilde t = 0$ and $\tilde t = \infty$ (provided $h(0,\infty)$ can be replaced with $\infty$, which is the case if $d h(0,\infty) \gg |d + i \tilde \Delta|^2$). Therefore, Eq.~(\ref{hequation}) can always be solved for $h(0,\tilde t)$. $|\tilde \Omega(\tilde t)|$ is then deduced by taking the square root of the derivative of $h(0,\tilde t)$. The phase of $\tilde \Omega$ is found by inserting $|\tilde \Omega|$ into Eq.~(\ref{etae2}) and is given by
\begin{eqnarray}\label{freeretphase}
\textrm{Arg}\left[\tilde \Omega(\tilde t)\right] & = & \pi + \textrm{Arg}\left[\eop_2(\tilde t)\right] - \frac{h(0,\tilde t)}{1+\tilde \Delta^2}\tilde \Delta
\nonumber \\ 
&&- \textrm{Arg}\Big[ \int_0^1 d \tilde z \frac{1}{1 + i \tilde \Delta} e^{-\frac{d \tilde z}{1 + i \tilde \Delta}}
\nonumber \\
 && \times I_0\left(2 \frac{\sqrt{h(0,\tilde t) d \tilde z}}{1+i \tilde \Delta}\right) S(1-\tilde z) \Big].  
\end{eqnarray}
The second and third terms are the phase of the desired output and the compensation for the Stark shift, respectively. In the resonant limit ($d \gamma \gg |\Delta|$), we can set $\tilde \Delta = 0$. Then assuming the phase of $S(\tilde z)$ is independent of $\tilde z$, the phase of the optimal control is given (up to a constant) solely by the phase of the desired output. In the Raman limit ($d \gamma \ll |\Delta|$), the integral in the last term is approximately real but at times near the end  of the output pulse ($t \approx T_\textrm{out}$) it can change sign and go through zero. At these times, the optimal $|\tilde \Omega(\tilde t)|$  diverges and the phase of $\tilde \Omega(\tilde t)$ changes by $\pi$. Numerical simulations show, however, that $|\tilde \Omega(\tilde t)|$ can be truncated at those points without significant loss in the retrieval efficiency and in the precision of $\eop_2(\tilde t)$ generation. Moreover, these points happen only near the back end of the desired output pulse over a rather short time interval compared to the duration of the desired output pulse. We can therefore often even completely turn off  $\tilde \Omega(\tilde t)$ during this short interval without significantly affecting the result, so that the problem of generating large power and $\pi$ phase shifts can be avoided altogether (see Sec.~\ref{sec:freeadcond} for another brief discussion of this issue). However, these potential difficulties in the Raman limit for generating the optimal control (which also has to be chirped according to Eq.~(\ref{freeretphase}) in order to compensate for the Stark shift) make the resonant (EIT) limit possibly more appealing than the far-off-resonant (Raman) limit.  

Finally, we note that a divergence in $|\tilde \Omega(\tilde t)|$ can occur at any detuning $\tilde \Delta$ even when the above Raman-limit divergences are not present. Specifically, similarly to the cavity discussion in paper I, if one wants to shape the retrieval into a mode $\eop_2(\tilde t)$ that drops to zero at some time $T_\textrm{out}$ sufficiently rapidly, $|\tilde \Omega(\tilde t)|$ will go to $\infty$ at $t = T_\textrm{out}$. However, as in the above case of the Raman-limit divergences, the infinite part can be truncated without significantly affecting the efficiency or the precision of $\eop_2(\tilde t)$ generation. One can confirm this by inserting into the adiabatic solution in Eq.~(\ref{freeadeout}) a control pulse that is truncated to have a value of $h(0,\infty)$ that is finite but large enough to satisfy $d h(0,\infty) \gg |d + i \tilde \Delta|^2$. However, to be completely certain that the truncation is harmless, one has to solve Eqs.~(\ref{freeeqs2e})-(\ref{freeeqs2s}) numerically without making the adiabatic approximation. We will do this in Sec.~\ref{sec:freeadcond} for the case of storage, where the same truncation issue is present. 

\subsection{Adiabatic storage \label{sec:freeadst}}

In principle, the retrieval results of the previous section and the time reversal argument immediately imply that in the adiabatic limit (see Sec.~\ref{sec:freeadcond} for precise conditions), any input mode $\eop_\textrm{in}(\tilde t)$ at any detuning $\tilde \Delta$ can be stored with the same $d$-dependent maximum efficiency if one appropriately shapes the control field. However, for completeness and to gain extra physical insight, in this section, we present an independent solution to adiabatic storage. 

Using the Laplace transform in space and a procedure similar to the one used in Sec.\ \ref{sec:freeadret} to solve retrieval, we find that the adiabatic solution of storage is
\begin{eqnarray} \label{freeadS}
S(\tilde z,\tilde T) &=&  - \sqrt{d} \int_0^{\tilde T} d \tilde t \tilde \Omega^*(t) \frac{1}{1 + i \tilde \Delta} e^{-\frac{h(\tilde t,\tilde T) + d \tilde z}{1 + i \tilde \Delta}} 
\nonumber \\ 
&&\times I_0\left( 2\frac{\sqrt{h(\tilde t,\tilde T) d \tilde z}}{1+i \tilde \Delta}\right) \eop_\textrm{in}(\tilde t).
\end{eqnarray}

It is important to note that the retrieval equation (\ref{freeadeout}) and the storage equation (\ref{freeadS}) can be cast in terms of the same $\Omega$-dependent function $m$ as
\begin{eqnarray} \label{timerevES}
\eop_\textrm{out}(\tilde t) &=& \int_0^1 d \tilde z \; m\!\!\left[\Omega(\tilde t'),\tilde t, \tilde z\right] S(1-\tilde z),
\\ \label{timerevSE}
S(\tilde z, \tilde T) &=& \int_0^{\tilde T} d \tilde t \; m\!\!\left[\Omega^*(\tilde T - \tilde t'),\tilde T - \tilde t, \tilde z\right] \eop_\textrm{in}(\tilde t). 
\end{eqnarray}
This is precisely the general time reversal property of our equations that we discussed abstractly in Secs.~\ref{sec:timerev} and \ref{sec:timerev2} and used to find the optimal storage strategy from optimal retrieval. However, as we said in the beginning of this section, we will now, for completeness, optimize storage directly without using time reversal and our solution for optimal retrieval. 

We would like to solve the following problem: given $\eop_\textrm{in}(\tilde t)$, $\Delta$, and $d$, we are interested in finding $\tilde \Omega(\tilde t)$ that will give the maximum storage efficiency. To proceed towards this goal, we note that if we ignore decay $\gamma$ and allow the spin wave to extend beyond $\tilde z = 1$, we get the ``decayless" storage equation
\begin{equation} \label{freeads}
s(\tilde z) =  \int_0^{\tilde T} d \tilde t q(\tilde z,\tilde t) \eop_\textrm{in}(\tilde t),
\end{equation}
where the ``decayless" mode $s(\tilde z)$ is defined for $\tilde z$ from $0$ to $\infty$ instead of from $0$ to $1$ and where
\begin{equation} \label{qztdef}
q(\tilde z,\tilde t) = i \sqrt{d} \tilde \Omega^*(\tilde t) \frac{1}{\tilde \Delta} e^{i \frac{h(\tilde t,\tilde T)+d \tilde z}{\tilde \Delta}} J_0\left( 2\frac{\sqrt{h(\tilde t,\tilde T) d \tilde z}}{\tilde \Delta}\right).
\end{equation}
Since in Eq.~(\ref{freeads}), both sources of storage loss (the decay $\gamma$ and the leakage past $\tilde z = 1$) are eliminated, the transformation between input modes $\eop_\textrm{in}(\tilde t)$ and decayless modes $s(\tilde z)$ becomes unitary. Indeed, we show in Appendix \ref{sec:appstor} that Eq.~(\ref{freeads}) establishes, for a given $\tilde \Omega(\tilde t)$, a 1-to-1 correspondence between input modes $\eop_\textrm{in}(\tilde t)$ and decayless modes $s(\tilde z)$. Moreover, we show in Appendix \ref{sec:appstor} that Eq.~(\ref{freeads}) also establishes for a given $s(\tilde z)$ a 1-to-1 correspondence between input modes $\eop_\textrm{in}(\tilde t)$ and control fields $\tilde \Omega(\tilde t)$. In particular, this means that we can compute the control field that realizes decayless storage (via Eq.~(\ref{freeads})) of any given input mode $\eop_\textrm{in}(\tilde t)$ into any given decayless spin-wave mode $s(\tilde z)$.   

A key element in the control shaping procedure just described is the ability to reduce the problem to the unitary mapping by considering the decayless (and leakless) solution. The reason why this shaping is useful and why it, in fact, allows us to solve the actual shaping problem in the presence of decay and leakage is that the spin wave, into which we store in the presence of decay, can be directly determined from the decayless solution: using Eqs.~(\ref{freeadS}) and (\ref{freeads}) and Eq.~(\ref{besselequality}) (with appropriate complex values of $\mu$, $\lambda$, and $p$), we find that
\begin{equation} \label{adddecay}
S(\tilde z,\tilde T) = \int_0^\infty d \tilde z' d e^{- d (\tilde z + \tilde z')} I_0(2 d \sqrt{\tilde z \tilde z'}) s(\tilde z').
\end{equation}
This means that, remarkably, $S(\tilde z,\tilde T)$ (and hence the storage efficiency) depends on $\eop_\textrm{in}(\tilde t)$ and $\tilde \Omega(\tilde t)$ only through the decayless mode $s(\tilde z)$, which itself can be computed via unitary evolution in Eq.~(\ref{freeads}). 

Computing storage efficiency from Eq.~(\ref{adddecay}) as a functional of $s(\tilde z)$ and maximizing it under the constraint that $s(\tilde z)$ is normalized gives an eigenvalue problem similar to Eq.~(\ref{eigeneqret}) except the upper limit of integration is $\infty$ and the kernel is different. After finding the optimal $s(\tilde z)$ via the iteration scheme similar to the one used to solve Eq.~(\ref{eigeneqret}), we conclude the procedure for optimal storage control shaping by using the unitary transformation in Eq.~(\ref{freeads}) to solve for the control in terms of $s(\tilde z)$ and $\eop_\textrm{in}(\tilde t)$, as shown in Appendix \ref{sec:appstor}. Since efficiency is determined by $s(\tilde z)$ alone, this gives the optimal storage with the same maximal efficiency for any input pulse shape (provided it is sufficiently smooth, as discussed in Sec.~\ref{sec:freeadcond}).  

Having derived the optimal storage control, we can now explicitly verify the results obtained from the time reversal reasoning. In Appendix \ref{sec:appstor}, we show that the mode $S(\tilde z,\tilde T)$ used in optimal storage is just the optimal mode for backward retrieval; that the optimal storage efficiency and optimal retrieval efficiency are equal; and that the optimal storage control for a given input mode is the time-reverse of the control that gives optimal backward retrieval into the time-reverse of that input mode.

\begin{figure}[tb]
\includegraphics[scale = 1]{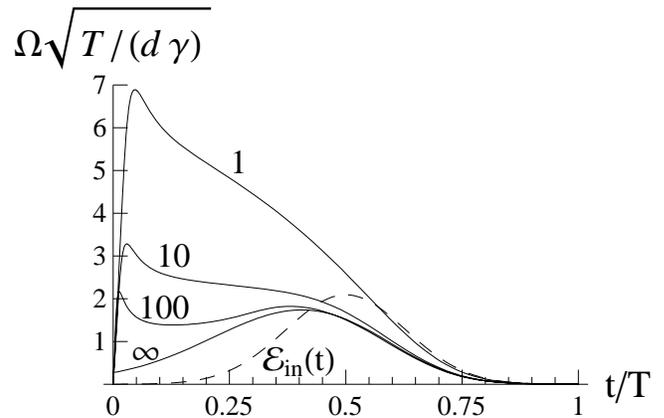}
\caption{Input mode $\eop_\textrm{in}(t)$ (dashed) defined in Eq.~(\ref{Gaussian}) and control fields $\Omega(t)$ (in units of $\sqrt{d \gamma/T}$) that maximize for this $\eop_\textrm{in}(t)$ the efficiency of resonant adiabatic storage (alone or followed by backward retrieval) at $d = 1, 10, 100$, and $d \rightarrow \infty$. \label{fig:modecontrols}}
\end{figure}

To give an example of optimal controls, we consider a Gaussian-like input mode (shown in Fig.~\ref{fig:modecontrols}) 
\begin{equation} \label{Gaussian}
\eop_\textrm{in}(\tilde t) = A(e^{-30 (\tilde t/\tilde T - 0.5)^2} - e^{-7.5})/\sqrt{\tilde T},
\end{equation} 
where for computational convenience we set $\eop_\textrm{in}(0) = \eop_\textrm{in}(\tilde T)=0$ and where $A \approx 2.09$ is a normalization constant. Figure \ref{fig:modecontrols} shows the corresponding optimal storage control shapes $\Omega$ for the case $\tilde \Delta = 0$ and $d = 1, 10, 100$, as well as the limiting shape of the optimal $\Omega$ as $d \rightarrow \infty$. The controls are plotted in rescaled units so that the area under the square of the curves shown is equal to $L^{-1} \int_0^T d t' v_\textrm{g}(t')$ (where $v_\textrm{g}$ is the EIT group velocity), which is the position (in units of $L$), at which the front end of the pulse would get stored under ideal decayless propagation. From time reversal and the condition for complete retrieval, it follows that the control pulse energy ($\propto h(0,\tilde T)$) and hence $L^{-1} \int_0^T d t' v_\textrm{g}(t')$ should diverge. Thus, at any finite $d$, the optimal $\Omega$ plotted in Fig.~\ref{fig:modecontrols} should actually diverge at $\tilde t = 0$. However, the front part of the control pulse affects a negligible piece of $\eop_\textrm{in}(\tilde t)$, so truncating this part (by truncating $s(\tilde z)$ at some $\tilde z$, for example) does not affect the efficiency (we will confirm this again in Sec.~\ref{sec:freeadcond}). Naively, the optimal control should roughly satisfy $L^{-1} \int_0^T d t' v_\textrm{g}(t') = 1$: the control (and, thus, the dark state polariton group velocity) should be small enough to avoid excessive leakage; and at the same time it should be as large as possible to have the widest possible EIT transparency window to minimize spontaneous emission losses. For the truncated optimal controls, we see that $L^{-1} \int_0^T d t' v_\textrm{g}(t')$ (i.e., the area under the square of the curves in Fig.~\ref{fig:modecontrols}) is, in fact, greater than $1$. As $d$ decreases, $L^{-1} \int_0^T d t' v_\textrm{g}(t')$ decreases as well, and allows for less and less leakage so that only as $d \rightarrow \infty$, $s(\tilde z) \rightarrow \sqrt{3}(1-\tilde z)$, $L^{-1} \int_0^T d t' v_\textrm{g}(t') \rightarrow 1$, and no leakage is allowed for.  

Similarly to the cavity storage discussed in paper I, although optimal storage efficiencies are the same in the Raman and adiabatic limits, the two limits exhibit rather different physical behavior. It is now the dependence on $d$ of the intensity of the optimal control field that can be used to distinguish between the resonant and the Raman regimes. Using an analysis very similar to pulse duration analysis of Sec.~\ref{sec:freeadret} or alternatively relying on the fact that optimal storage is the time-reverse of optimal retrieval, we find that in the resonant limit ($d \gamma \gg |\Delta|$),  $|\Omega|^2 \sim d \gamma/T$, while in the Raman limit ($d \gamma \ll |\Delta|$), $|\Omega|^2 \sim \Delta^2/(\gamma T d)$. Both of these agree with the cavity model of paper I (identifying $C$ and $d$) while the resonant control agrees with $v_\textrm{g} T \sim L$. As explained in the cavity case in Sec.\ V B of paper I, this opposite dependence of $|\Omega|$ on $d$ in the Raman and EIT limits is the consequence of the fact that the coupling of the input photon to the spin wave has the opposite dependence on $\Omega$ in the two regimes. Finally, we note that in the Raman limit ($|\Delta| \gg d \gamma$), $|\Omega|^2 \sim \Delta^2/(\gamma T d) \gg d \gamma/T$, which means that if one is limited by control power (as in Ref.~\cite{nunn06}), the EIT regime is preferable to the Raman regime.

\subsection{Storage followed by retrieval\label{sec:freeadstret}}

In the cavity case discussed in paper I, there was only one spin-wave mode we could write on. Moreover, this spatially uniform mode looked the same in the forward and backward directions (assuming negligible metastable state splitting $\omega_{sg}$). Therefore, optimal storage into that mode guaranteed that the combined procedure of storage followed by retrieval was optimal as well and the total efficiency did not depend on the retrieval direction. In contrast, the free-space model allows for storage into a variety of modes, each of which has a different retrieval efficiency that is also dependent on retrieval direction. Therefore, in free space, we will first discuss the optimization of storage followed by backward retrieval and then the optimization of storage followed by forward retrieval.

\begin{figure}[t]
\begin{center}
\includegraphics[scale = 1]{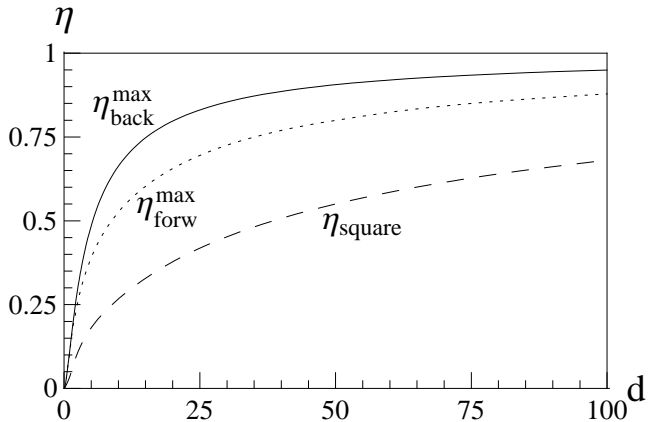}
\end{center}
\caption{$\eta^\textrm{max}_\textrm{back}$ (solid) and $\eta^\textrm{max}_\textrm{forw}$ (dotted) are maximum total efficiency for adiabatic (or fast) storage followed by backward or forward retrieval, respectively. $\eta_\textrm{square}$ (dashed) is the total efficiency for resonant storage of  $\eop_\textrm{in}(t)$ from Eq.~(\ref{Gaussian}) followed by backward retrieval, where the storage control field is a naive square pulse with the power set by $v_\textrm{g} T = L$. \label{fig:efficplota}}
\end{figure}

Since we have shown that the optimal spin-wave mode for backward retrieval is also the optimal mode for storage, the controls found in Sec.~\ref{sec:freeadst}, which optimize storage, are also optimal for storage followed by backward retrieval. Figure \ref{fig:efficplota} shows the maximum total efficiency for storage followed by backward retrieval ($\eta^\textrm{max}_\textrm{back}$ - solid line), which in the adiabatic limit can be achieved for any input pulse. For comparison, we also show the total efficiency for storage followed by backward retrieval for a Gaussian-like input mode defined in Eq.~(\ref{Gaussian}) (assuming the adiabatic limit $T d \gamma \gg 1$) with naive square storage control pulses on $[0,T]$  with power set by $v_\textrm{g} T = L$, where $v_\textrm{g}$ is the EIT group velocity ($\eta_\textrm{square}$ - dashed line).  The significant increase in the efficiency up to the input-independent optimal efficiency $\eta^\textrm{max}_\textrm{back}$ due to the use of optimal storage control pulses instead of naive ones is, of course, not unique to the input pulse of Eq.~(\ref{Gaussian}) and holds for any input pulse. In fact, since at moderate values of $d$ the naive control pulse is far from satisfying the complete retrieval condition, it is not optimal for any input.
  
Since the optimal mode for storage or retrieval alone is not symmetric, a separate optimization problem has to be solved for the case of storage followed by forward retrieval. We show in Appendix \ref{sec:appstor} that Eq.~(\ref{freeads}) sets up, for any sufficiently smooth $\eop_\textrm{in}(\tilde t)$, a $1$-to-$1$ correspondence between decayless modes $s(\tilde z)$ and control fields $\tilde \Omega(\tilde t)$. Moreover, the decayless mode alone determines the total efficiency of storage followed by forward retrieval, which can be found by inserting Eq.~(\ref{adddecay}) into Eq.~(\ref{freereteff}). Thus, the optimization problem reduces to finding the optimal $s(\tilde z)$ by the iterative optimization procedure except with a new kernel.

From a different perspective, since the process of storage followed by forward retrieval as a whole fits into the general setup of Sec.~\ref{sec:timerev2}, we can use time reversal to optimize it. In particular, we showed in Sec.~\ref{sec:timerev2} how to find the optimal input mode to be used with a given pair of control pulses: we should take a trial input mode, carry out storage followed by forward retrieval, time reverse the output and the controls, and iterate until convergence. Since the reverse process is itself storage followed by forward retrieval (except in the opposite direction) and since the iterations will optimize it as well, the spin-wave mode used in optimal storage followed by forward retrieval must be the one that flips under forward retrieval, followed by time reversal and (backward) storage. Moreover, it follows that the control pulse that we should use for a given $\eop_\textrm{in}(\tilde t)$ is the time-reverse of the control that retrieves the flipped version of the optimal spin-wave mode backwards into $\eop^*_\textrm{in}(\tilde T-\tilde t)$.  

Thus, instead of computing the optimal $s(\tilde z)$, we can solve the following eigenvalue problem that finds the optimal mode to store into
\begin{equation} \label{forwardoptim}
\lambda S(\tilde z) = \int_0^1 d \tilde z' k_\textrm{r}(\tilde z,1-\tilde z') S(\tilde z').
\end{equation}
This eigenvalue equation is just a simple modification of the retrieval eigenvalue equation (\ref{eigeneqret}): we are now computing the mode that flips under forward retrieval followed by time reversal and storage, while Eq.~(\ref{eigeneqret}) finds the mode that stays the same under backward retrieval followed by time reversal and storage. The total efficiency of storage followed by forward retrieval is then $\lambda^2$. However, in contrast to storage followed by backward retrieval, the storage efficiency and the forward retrieval efficiency during the optimal procedure are not each equal to $\lambda$; the storage efficiency is greater.

\begin{figure}[tb]
\begin{center}
\includegraphics[scale = 1]{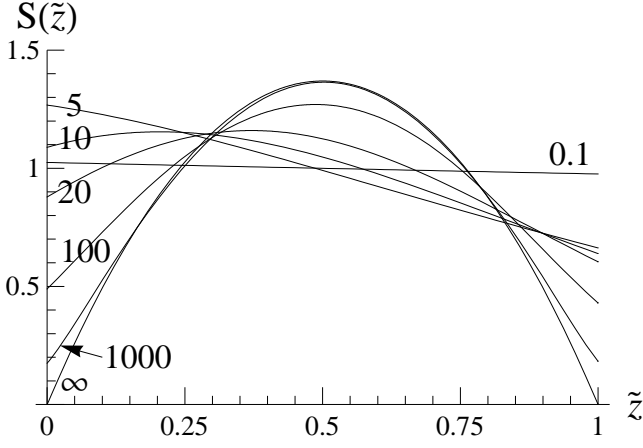}
\end{center}
\caption{For different values of $d$, the optimal spin-wave mode to be used during storage followed by forward retrieval. \label{fig:forwardmodes}}
\end{figure}

The optimal spin-wave modes that result from Eq.~(\ref{forwardoptim}) are shown in Fig.~\ref{fig:forwardmodes} for the indicated values of $d$. At small $d$, the optimal mode is almost flat since at small $d$ the optimal retrieval and storage modes are almost flat and, thus, almost symmetric. As $d$ increases, the optimal mode first bends towards the wedge $\sqrt{3} (1-\tilde z)$ similarly to the optimal backward retrieval mode. But then above $d \approx 3$, the optimal mode starts shaping towards the parabola $S(z)=\sqrt{15/8}(1 - 4(\tilde z - 1/2)^2)$, which, as expected, avoids the $1/\sqrt{d}$ error from discontinuities by vanishing at the edges and, simultaneously, maximizes smoothness.    

The maximal total efficiency for storage followed by forward retrieval $\eta^\textrm{max}_\textrm{forw}$ is plotted as a dotted curve in Fig.~\ref{fig:efficplota}. $\eta^\textrm{max}_\textrm{forw}$ ($\approx 1 - 19/d$ as $d \rightarrow \infty$) is less than $\eta^\textrm{max}_\textrm{back}$ ($\approx 1 - 5.8/d$ as $d \rightarrow \infty$) since for optimal backward retrieval, storage and retrieval are each separately optimal, while for forward retrieval a compromise has to be made. From a different perspective, forward retrieval makes it more difficult to minimize propagation since the whole excitation has to propagate through the entire medium. 

\subsection{Adiabaticity conditions \label{sec:freeadcond}}
 
We have shown that in the adiabatic limit, any input mode can be stored with the same maximum efficiency. 
In this section, we show that independent of $\Delta$, the necessary and sufficient condition for optimal adiabatic storage of a pulse of duration $T$ to be consistent with the adiabatic approximation is 
\begin{equation} \label{freecond}
T d \gamma \gg 1,
\end{equation}
which is identical to the corresponding condition in the cavity model in paper I, except with $C$ replaced with $d$. In fact, we omit the derivation of this condition since the argument goes along the same lines as the corresponding argument in the cavity case in Sec.~V C of paper I, provided one uses the fact that in Eq.~(\ref{freeeqs2pl}) of the present paper $|u| \sim 1$ (since $u$ is the Laplace variable corresponding to $\tilde z$, which, in turn, runs from $0$ to $1$).

\begin{figure}[t]
\begin{center}
\includegraphics[scale = 1]{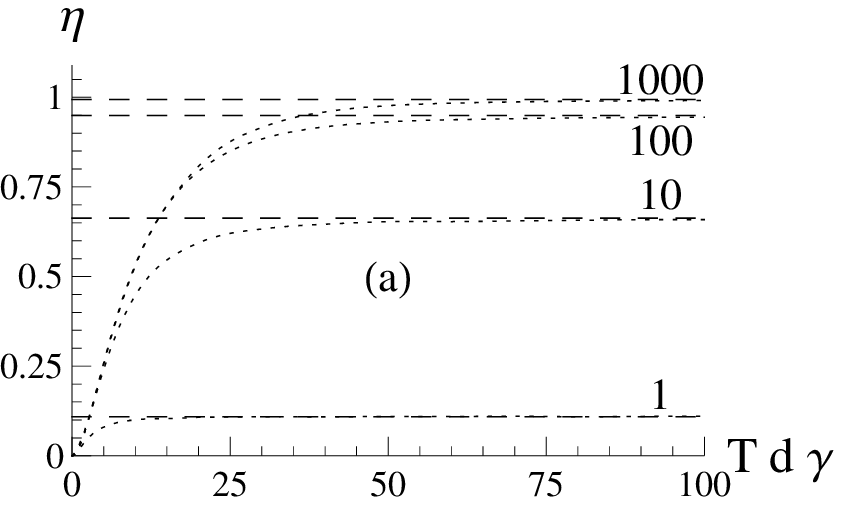}
\includegraphics[scale = 1]{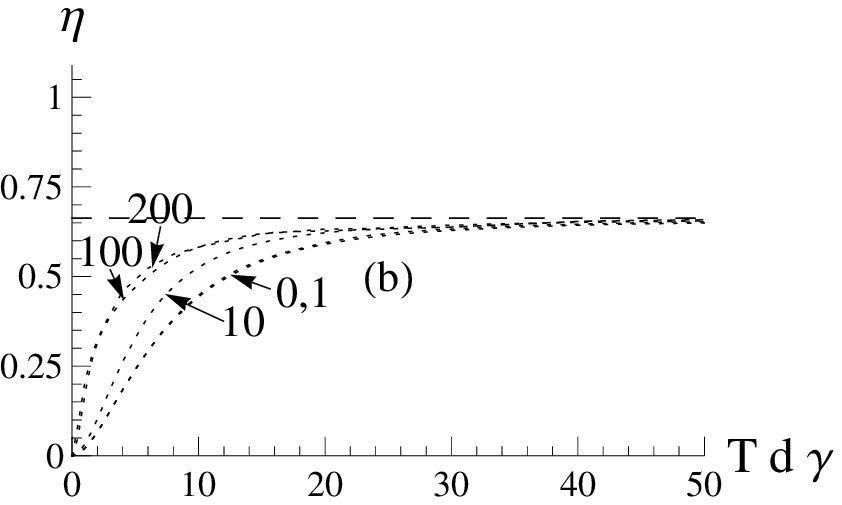}
\end{center}
\caption{Breakdown of optimal adiabatic storage in free space at $T d \gamma \lesssim 10$. In (a), the total efficiency of storage followed by backward retrieval is plotted for $\Delta = 0$ and $d = 1, 10, 100$, and $1000$. The horizontal dashed lines are the maximal values. Dotted lines are obtained for the input from Fig.~\ref{fig:modecontrols} using adiabatic equations to shape the storage control but then using exact equations to numerically compute the efficiency. In (b), the same plot is made for $d = 10$ and $\Delta/\gamma = 0, 1, 10, 100, 200$.   \label{fig:freeadbreak}}
\end{figure}

Therefore, we immediately turn to the tasks of verifying numerically that Eq.\ (\ref{freecond}) is indeed the correct adiabaticity condition and of investigating the breakdown of adiabaticity for short input pulses. We consider adiabatic storage of a Gaussian-like input mode defined in Eq.~(\ref{Gaussian}) and shown in Fig.~\ref{fig:modecontrols}. We use our adiabatic equations to shape the control pulse but then compute the total efficiency of storage followed by backward retrieval numerically  from Eqs.~(\ref{freeeqs2e})-(\ref{freeeqs2s}) \textit{without} making the adiabatic approximation. As $T d \gamma$ decreases to $1$ and below, we expect the efficiency to be reduced from its optimal value. In Fig.~\ref{fig:freeadbreak}(a), the total efficiency of this procedure is plotted as a function of $T d \gamma$ for $\Delta=0$ and $d = 1, 10, 100$, and $1000$. The dashed lines are the true optimal efficiencies. As expected, when $T d \gamma \lesssim 10$, the efficiency drops. In Fig.~\ref{fig:freeadbreak}(b), we fix $d = 10$ and show how optimal adiabatic storage breaks down at different detunings $\Delta$ from $0$ to $200 \gamma$. As in the cavity case of paper I, we see from Fig.~\ref{fig:freeadbreak}(b) that as we move from the resonant limit ($d \gamma \gg |\Delta|$) to the Raman limit ($d \gamma \ll |\Delta|$), we can go to slightly smaller values of $T d \gamma$ before storage breaks down. However, since the curves for $\Delta = 100 \gamma$ and $\Delta = 200 \gamma$ almost coincide, $T d \gamma \gg 1$ is still the relevant condition no matter how high $\Delta$ is.   

Before concluding the discussion of adiabaticity conditions, we note that, exactly as in the cavity case discussion in Sec.\ V C of paper I, the magnitudes $|\Omega(t)|$ of the optimal control pulses, which go to $\infty$ at $t = 0$, had to be trucated to generate Fig.~\ref{fig:freeadbreak}. Moreover, Raman-limit divergences discussed (for the case of retrieval) in Sec.~\ref{sec:retshaping} had to be truncated at two places near $t = 0.2 \, T$ for $\Delta/\gamma = 100$ and $200$ in Fig.~\ref{fig:freeadbreak}(b). As in Sec.\ V C of paper I, the fact that the optimal efficiency given by the dashed lines in Fig.~\ref{fig:freeadbreak} is achieved by the dotted curves (obtained with truncated controls) is the proof that truncation of the storage controls does not significantly affect the storage efficiency. Since time reversal discussed Sec.~\ref{sec:timerev} relates retrieval to optimal storage, this also means that truncation of the retrieval control fields does not significantly affect the precision, with which a given output mode $\eop_2(\tilde t)$ can be generated. The losses associated with truncation are insignificant only if condition $d h(0,\infty) \gg |d + i \tilde \Delta|^2$ (see Sec.~\ref{sec:freeadretsol}) is satisfied for the truncated retrieval control pulse (the same condition with $\infty$ replaced with $\tilde T$ applies to storage). If the energy in the control pulse is so tightly limited that this condition is violated, a separate optimization problem has to be solved. This problem has been considered in Ref.~\cite{nunn06} for the special case of Raman storage in free space in the limit of negligible spontaneous emission loss. Although Eqs.~(\ref{freeadeout}) and (\ref{freeadS}) allow one to treat and optimize the case of limited control pulse energy, this problem is beyond the scope of the present paper.

\subsection{Effects of nonzero spin-wave decay \label{sec:freespindecay}}

We have so far assumed that the spin-wave decay rate in Eq.~(\ref{freeeqss}) is negligible (i.e.,  $\gamma_\textrm{s} = 0$). If this decay rate is not negligible but the corresponding incoming noise is vacuum (which is often true experimentally, as noted in Appendix \ref{sec:appModelfree} of the present paper and as explained in detail in Appendix A of paper I), the effect of nonzero spin-wave decay can be included simply by adding a term $-\tilde \gamma_\textrm{s} S$ to the right-hand side of Eq.~(\ref{freeeqs2s}) (where $\tilde \gamma_\textrm{s} = \gamma_\textrm{s}/\gamma$). In this section, we discuss the effects such a term has on the adiabatic solution and optimal control field shaping discussed in Secs.~\ref{sec:freeadret},  \ref{sec:freeadst}, and \ref{sec:freeadstret}. We will show, in particular, that nonzero $\gamma_\textrm{s}$ simply introduces exponential decay into the retrieval and storage solutions without making the optimization harder. We will also show that with nonzero spin-wave decay, optimal efficiencies become dependent on the input mode (or, equivalently, on the control fields).   

We first consider adiabatic retrieval discussed in Sec.~\ref{sec:freeadret}. One can easily check that nonzero spin-wave decay rate $\gamma_\textrm{s}$ simply introduces decay described by $\exp(- \tilde   \gamma_\textrm{s} \tilde t)$ into Eq.~(\ref{freeadeout}), and, unless we retrieve much faster than $1/\gamma_\textrm{s}$, this makes the retrieval efficiency control dependent. Moreover, if a given fixed control field is not strong enough to accomplish retrieval in a time much faster than $1/\gamma_\textrm{s}$, the problem of finding the optimal retrieval mode for this particular retrieval control will give a different answer from the $\gamma_\textrm{s} = 0$ case. In particular, for forward retrieval, as we increase $\gamma_\textrm{s}$ (or, alternatively, decrease the power of the retrieval control), to minimize propagation time at the cost of sacrificing smoothness, the optimal retrieval mode will be more and more concentrated towards the $z = L$ end of the ensemble. As in the $\gamma_\textrm{s} = 0$ case, we can find these optimal modes either by computing the (now $\Omega$-dependent) kernel to replace $k_\textrm{r}$ in Eq.~(\ref{freereteff}) and its eigenvector with the largest eigenvalue or, equivalently, by doing the iteration of retrieval, time reversal, and storage.

The inclusion of nonzero $\gamma_\textrm{s}$ also does not prevent us from being able to shape retrieval to go into any mode, as described for $\gamma_\textrm{s}=0$ in Sec.~\ref{sec:retshaping}. We should just shape the control according to Eq.~(\ref{etae2}) as if there were no spin wave decay except the desired output mode $\eop_2(t)$ on the left hand side should be replaced with the normalized version of $\eop_2(t) \exp(\gamma_\textrm{s} t)$, i.e.,
\begin{equation}
\eop_2(\tilde t) \rightarrow \eop_2(\tilde t) e^{\tilde \gamma_\textrm{s} \tilde t} \left[\int_0^\infty d \tilde t' |\eop_2(\tilde t')|^2 e^{2 \tilde \gamma_\textrm{s} \tilde t'}\right]^{-\frac{1}{2}}.
\end{equation}
The retrieval efficiency will, however, be output-mode-dependent in this case: it will be multiplied (and hence reduced) by a factor of $\left[\int_0^\infty d \tilde t' |\eop_2(\tilde t')|^2 \exp(2 \tilde \gamma_\textrm{s} \tilde t')\right]^{-1}$. Since this factor is independent of the spin wave, even with nonzero $\gamma_\textrm{s}$, the optimal retrieval into $\eop_2(t)$ is achieved by retrieving from $\tilde S_d(z)$ .

We now turn to adiabatic storage discussed in Sec.~\ref{sec:freeadst}. One can easily check that nonzero spin-wave decay $\gamma_\textrm{s}$ simply introduces $\exp(-\tilde \gamma_\textrm{s}(\tilde T-\tilde t))$ decay into Eq.~(\ref{qztdef}) (or, equivalently, into the integrand on the right hand side of Eq.~(\ref{freeadS})). Eq.~(\ref{adddecay}) holds even with nonzero $\gamma_\textrm{s}$. The optimal storage control can then be found from Eq.~(\ref{freeads}) as if there were no decay but the input mode were replaced according to
\begin{equation}\label{rescaled}
\eop_\textrm{in}(\tilde t) \rightarrow \eop_\textrm{in}(\tilde t) e^{-\tilde \gamma_\textrm{s}(\tilde T-\tilde t)} \left[\int_0^{\tilde T}\!\! d \tilde t' |\eop_\textrm{in}(\tilde t')|^2 e^{-2 \tilde \gamma_\textrm{s}(\tilde T-\tilde t')}\right]^{-\frac{1}{2}}.
\end{equation}
However, the optimal storage efficiency will now depend on input pulse duration and shape: it will be multiplied (and hence reduced) by $\int_0^{\tilde T} d \tilde t' |\eop_\textrm{in}(\tilde t')|^2 \exp(-2 \tilde \gamma_\textrm{s}(\tilde T-\tilde t'))$. It is also important to note that nonzero $\gamma_\textrm{s}$ still keeps the general time reversal relationship between storage and retrieval exhibited in Eqs.~(\ref{timerevES}) and (\ref{timerevSE}). However, for $\eop_2(\tilde t) = \eop^*_\textrm{in}(\tilde T - \tilde t)$,
\begin{equation}
\int_0^{\tilde T} d \tilde t |\eop_\textrm{in}(\tilde t)|^2 e^{-2 \tilde \gamma_\textrm{s}(\tilde T-\tilde t)} > \left[\int_0^{\tilde T} d \tilde t |\eop_2(\tilde t)|^2 e^{2 \tilde \gamma_\textrm{s} \tilde t}\right]^{-1}, 
\end{equation}
which means that with nonzero $\gamma_\textrm{s}$, the optimal storage efficiency of a given input mode is greater than the optimal retrieval efficiency into the time-reverse of that mode. Because the two controls involved are not time-reverses of each other, the inequality of the two efficiencies is consistent with the time reversal arguments. As in the cavity case in paper I, the main reason for this deviation from the $\gamma_\textrm{s} = 0$ behavior is the dependence of the retrieval efficiency on the control. 

Finally, we discuss the effects of nonnegligible spin-wave decay on storage followed by retrieval considered in Sec.~\ref{sec:freeadstret}. Using the fact that nonzero $\gamma_\textrm{s}$ keeps the general time reversal relationship between storage and retrieval exhibited in Eqs.~(\ref{timerevES}) and (\ref{timerevSE}), it is not hard to verify that nonzero $\gamma_\textrm{s}$ still allows one to use time reversal iterations to optimize storage followed by retrieval. In particular, suppose that one is given a storage control field and a (forward or backward) retrieval control field. Then one can find the optimal input mode to be used with these control fields by the following procedure: start with a trial input mode, store and retrieve it with the given pair of controls, time-reverse the whole procedure, and then repeat the full cycle until convergence is reached. Now suppose, on the other hand, one is given an input mode and is asked to choose the optimal storage and retrieval controls. Because of the spin wave decay, it is desirable to read out as fast as possible. As we discuss in the next section, fast readout may be achieved in a time $T\sim 1/\gamma d$, so that if we assume that $\gamma_\textrm{s} \ll d\gamma$, the spin-wave decay during the retrieval will be negligible. If we further assume that the given input mode satisfies the adiabatic limit $T d \gamma \gg 1$, then one should shape the storage control to store into the appropriate optimal spin-wave mode ($\tilde S_d(1-\tilde z)$ or a mode from Fig.~\ref{fig:forwardmodes}, depending on the direction of retrieval) as if $\gamma_\textrm{s}$ were zero and the input were proportional to $\eop_\textrm{in}(\tilde t) \exp(-\tilde \gamma_\textrm{s}(\tilde T-\tilde t))$ (see Eq.~(\ref{rescaled})). The total optimal efficiency will now depend on input pulse duration and shape: it will be multiplied (and hence reduced relative to the $\gamma_\textrm{s}=0$ case) by $\int_0^{\tilde T} d \tilde t |\eop_\textrm{in}(\tilde t)|^2 \exp(-2 \tilde \gamma_\textrm{s}(\tilde T-\tilde t))$. Finally, we note that when we consider storage followed by retrieval, in order to take into account the spin wave decay during the storage time $[\tilde T, \tilde T_\textrm{r}]$, one should just multiply the total efficiency by $\exp(-2 \tilde \gamma_\textrm{s} (\tilde T_\textrm{r}-\tilde T))$.

\section{Fast Retrieval and Storage \label{sec:freefast}}

We have shown that in the adiabatic limit ($T d \gamma \gg 1$, where $T$ is the duration of the incoming pulse), one can optimally store a mode with any smooth shape and any detuning $\Delta$. In this section, we solve Eq.~(\ref{freeoneeq}) analytically in the second important limit, the ``fast" limit, and demonstrate that this limit allows one to store optimally a certain class of input modes that have duration $T \sim 1/(d \gamma)$. We also show that efficient (but not optimal) fast storage of any smooth pulse is possible as long as $T \gamma \ll 1$ and $T d \gamma \gg 1$. 

Exactly as in the cavity case in Sec.\ VI of paper I, in the fast limit, one assumes that $\Omega$ is very large during a short control pulse ($|\Omega| \gg d \gamma$ and $|\Omega| \gg |\Delta|$) and keeps only terms containing $\tilde \Omega$ on the right-hand side of Eqs.~(\ref{freeeqs2p}) and (\ref{freeeqs2s}) (or, equivalently, neglects all terms in  Eq.~(\ref{freeoneeq}) except $|\tilde \Omega|^2 S$ and $\ddot{S}$).  This gives Rabi oscillations between $P$ and $S$ and allows one to implement a fast storage scheme, in which the input pulse is resonant ($\Delta = 0$) and the control pulse is a short $\pi$ pulse at $t = T$, as well as fast retrieval, in which the control is a $\pi$ pulse at $t=T_\textrm{r}$.

During fast retrieval, assuming that the $\pi$ pulse is perfect and that it enters the medium at $\tilde t = 0$ (instead of $\tilde t = \tilde T_\textrm{r}$), the initial spin wave $S = S(\tilde z)$ is mapped after the $\pi$ pulse onto the optical polarization $P = i S(\tilde z)$. We then solve Eq.~(\ref{freeeqs2pl}) for $P(u,\tilde t)$, express $\eop(u, \tilde t)$ in terms of $P(u,\tilde t)$ using Eq.~(\ref{freeeqs2el}), and take the inverse Laplace transform $u \rightarrow \tilde z = 1$ to arrive at 
\begin{equation} \label{freefasteout}
\eop_\textrm{out}(\tilde t) = - \sqrt{d} \int_0^1 d \tilde z e^{-\tilde t} J_0\left(2 \sqrt{d \tilde t \tilde z}\right) S(1-\tilde z).
\end{equation}
When computing the fast retrieval efficiency, one can take the time integral analytically to find that the efficiency is again given by Eq.~(\ref{etargen}), which is consistent with the general proof in Sec.~\ref{sec:freeret} and the branching ratio argument. In the cavity case discussion in paper I, we noted that the fast solution was a special case of the adiabatic solution with a suitable control. Similarly, the expression in Eq.~(\ref{freefasteout}) is also a special case of Eq.~(\ref{freeadeout}) if we use
\begin{equation} \label{fastom}
\tilde \Omega(\tilde t) = (1 + i \tilde \Delta) e^{-i \tilde \Delta \tilde t}
\end{equation} 
and take the limit $\tilde \Delta \rightarrow \infty$ (although this violates the approximations made in deriving Eq.~(\ref{freeadeout})).

Since the $\pi$-pulse control field in fast retrieval is fixed, optimal fast retrieval yields a single possible output mode, that of Eq.~(\ref{freefasteout}) with the optimal spin wave $S(\tilde z) = \tilde S_d(\tilde z)$. By time reversal, the time-reversed version of this input mode (of duration $T \sim 1/(\gamma d)$) is, therefore, the only mode that can be optimally stored using fast storage at this optical depth $d$.
  
In order to confirm the time reversal argument and for the sake of completeness, one can also compute the optimal input mode for fast storage directly. For an input mode $\eop_\textrm{in}(\tilde t)$ nonzero for $\tilde t \in [0, \tilde T]$ and assuming that a perfect $\pi$ pulse arrives at $\tilde t= \tilde T$, we find that
\begin{equation} \label{freefastst}
S(\tilde z,\tilde T) \!=\! -\sqrt{d} \int_0^{\tilde T}\!\!\! d \tilde t e^{-(\tilde T-\tilde t)} J_0\!\left(\!2 \sqrt{d (\tilde T-\tilde t) \tilde z}\right) \!\eop_\textrm{in}(\tilde t).
\end{equation}
One can see that the fast retrieval and storage equations (\ref{freefasteout}) and (\ref{freefastst}) obey, as expected, the same general time reversal relationship that we have already verified in the adiabatic limit in Eqs.~(\ref{timerevES}) and (\ref{timerevSE}). One can also explicitly verify that the maximization of the storage efficiency derived from Eq.~(\ref{freefastst}) yields an optimal $\eop_\textrm{in}(\tilde t)$ that is the normalized time reverse of Eq.~(\ref{freefasteout}) evaluated with the optimal spin wave $S(\tilde z) = \tilde S_d(\tilde z)$. It is worth noting that short exponentially varying pulses, reminiscent of our optimal solution, have been proposed before to achieve efficient photon-echo based storage \cite{kalachev}.

The solutions above give an incoming mode $\eop_{\textrm{in}}(\tilde t)$ that is optimal for fast storage alone or for fast storage followed by backward retrieval. Similarly, at each $d$, there is a mode that gives the optimal efficiency for fast storage followed by forward retrieval. This optimal input mode is the time-reverse of the output of fast forward retrieval from the spin-wave mode optimal for storage followed by forward retrieval (as computed through Eq.~(\ref{forwardoptim}) and shown in Fig.~\ref{fig:forwardmodes}).

Finally, we note an important difference between fast storage in a cavity discussed in paper I and fast storage in free space. In a cavity, there is only one accessible spin-wave mode, and hence only one input mode that can be stored using fast storage (i.e., any input mode orthogonal to it will have zero storage efficiency). As shown in paper I, this input mode is exponentially rising with a time constant $\sim 1/(\gamma C)$, where $C$ is the cooperativity parameter. Therefore, generating this mode, and hence obtaining high efficiency, may be hard in practice at high values of $C$. In contrast, in free space, any sufficiently smooth spin wave will have a high retrieval efficiency, and, by time-reversal, the time-reverses of the pulses fast retrieved from these spin waves can also be fast stored with high efficiency. One can, thus, explicitly verify using Eq.~(\ref{freefastst}), which allows one to compute these storage efficiencies, that if, in the original units, $T \gamma \ll 1$ but at the same time $T d \gamma \gg 1$, the free-space fast-storage efficiency is close to unity.

\section{Effects of Metastable State Nondegeneracy \label{sec:nondeg}}

In the discussion of backward retrieval we have so far assumed that the two metastable states $|g\rangle$ and $|s\rangle$ are nearly degenerate. This has meant that during backward retrieval we could simply use the same equations as for forward retrieval but with the spin wave flipped: $S(z)\rightarrow S(L-z)$. If $|g\rangle$ and $|s\rangle$ are not degenerate and are split by $\omega_{sg} = c \Delta k$, then during backward retrieval, instead of retrieving from $S(L- z)$, we will have to redefine the slowly varying operators (see Eq.~(\ref{slowops4})) and retrieve from $S(L-z) \exp(-2 i \Delta k z)$, which significantly lowers the efficiency unless $\Delta k L \ll \sqrt{d}$. This condition on $\Delta k$ can be understood based on the concept of the effective EIT window for the Fourier transform of the spin wave. As explained in Sec.~\ref{sec:retddep}, the width of this window is of order (in the original units) $\sim \sqrt{d}/L$. The extra phase just shifts the Fourier transform off center by $2 \Delta k$, so that the efficiency will not be significantly affected provided the shift is much smaller than the window width. We have  confirmed numerically for $S(\tilde z) = 1$ and for $S(\tilde z) = \sqrt{3} \tilde z$ that the $\Delta k L$ needed to decrease retrieval efficiency by $50\%$ from its $\Delta k L = 0$ value indeed scales as $\sqrt{d}$ (with proportionality constants $\approx 0.46$ and $\approx 0.67$, respectively).
 
There are two ways to understand physically why nondegeneracy of the metastable states ruins the backward retrieval efficiency. The first explanation, also noted in Ref.~\cite{nunn06a}, comes from the fact that metastable state nondegeneracy breaks the momentum conservation on backward retrieval. During storage, momentum $\Delta k$ is written onto the ensemble. Momentum conservation on backward retrieval, however, will require $-\Delta k$ momentum in the spin wave. The second explanation comes from the fact that if $\Delta k \neq 0$, then backward retrieval of optimal storage is no longer its time reverse. If we had not defined slowly varying operators, the spin wave that we store into our atoms would have had $\exp(i \Delta k z)$ phase written on it. Since time reversal consists of moving in the opposite direction and taking a complex conjugate, backward retrieval will be the time-reverse of storage only if $\Delta k = 0$, in which case complex conjugation is trivial. Thus, if $\Delta k \neq 0$, the optimization of storage does not simultaneously optimize backward retrieval (unless, of course, we can apply the desired position-dependent phase to the atoms during the storage time, e.g., by a magnetic field gradient, or alternatively apply a $\pi$ pulse that flips the two metastable states \cite{nunn06a}).  

We would like now to optimize storage followed by backward retrieval in the presence of nondegeneracy ($\Delta k \neq 0$). Following the general recipe of Sec.~\ref{sec:timerev2}, in order to carry out the optimization, one has to start with an input pulse and a control pulse, do storage, then do backward retrieval with another control pulse. Then one has to time reverse the full process of storage and retrieval, and iterate till one gets convergence to a particular input (and spin wave). Specifically, we start with a trial spin wave $S_1(z)$. To find the spin wave (in terms of operators that are slowly varying for forward propagation as defined in Eq.~(\ref{slowops4})) that the optimal storage plus backward retrieval should use, we first rewrite $S_1(z)$ for backward-propagation slowly varying operators (i.e., add the $2 \Delta k z$ phase), and then retrieve it backwards, time reverse, and store. Using Eq.~(\ref{generalizedeigeneqret}), the iteration we get is (dropping an unimportant constant phase and going to our rescaled units) 
\begin{equation} \label{deltakiter}
S_2(\tilde z) = \int_0^1 d\tilde{z}' k_\textrm{r}(\tilde z,\tilde z') e^{- i 2 \Delta \tilde k \tilde z'} S_1^*(\tilde z'), 
\end{equation}
where $\Delta \tilde k = L \Delta k$. This iteration finds the eigenvector with the largest eigenvalue for the eigenvalue problem 
\begin{equation} 
\lambda S(\tilde z) = \int_0^1 k_\textrm{r}(\tilde z,\tilde z') e^{- i 2 \Delta \tilde k \tilde z'} S^*(\tilde z').
\end{equation}
$|\lambda|^2$ will then give the total maximum efficiency of storage followed by backward retrieval. In contrast to the $\Delta k = 0$ case, the efficiencies of storage and retrieval in the optimal process are not generally equal. It is important to note that since the process we are optimizing followed by its time reverse corresponds to two iterations of Eq.~(\ref{deltakiter}), after a sufficient number of steps, $\lambda$ settles into an oscillation between $|\lambda| \exp(i \alpha)$ and $|\lambda| \exp(-i \alpha)$ for some phase $\alpha$. The eigenvector will oscillate between two values differing only by an unimportant constant phase, so that either one can be used.

While this procedure allows us to find the optimal spin waves, we should, for completeness, also determine, as in the $\Delta k = 0$ case, which input fields the optimum can be achieved for. To do this, we, as before, consider the exactly solvable adiabatic and fast limits.  
In the adiabatic limit, the argument that retrieval can be shaped into any mode did not require the spin wave $S(z)$ to be real, and it is therefore still applicable.  
By time reversal, we can, therefore, still achieve the maximum efficiency of storage followed by backward retrieval for any incoming mode of duration $T$ such that $T d \gamma \gg 1$.  Similarly, in the fast limit, using fast retrieval and time reversal we can find at each $d$ a pulse shape with $T d \gamma \sim 1$ that gives the maximum efficiency. For completeness, we note that one can also generalize to $\Delta k \neq 0$ the method that uses the decayless mode $s(z)$ to shape the optimal storage control, as described in Sec.~\ref{sec:freeadst}. However, since the optimal control is unique, this method will, of course, yield the same control as the method based on retrieval and time reversal (as we showed explicitly in Sec.~\ref{sec:freeadst} and Appendix \ref{sec:appstor} for $\Delta k = 0$). We will, thus, omit here the extension of this method to the $\Delta k \neq 0$ case.

\begin{figure}[tb]
\begin{center}
\includegraphics[scale = 1]{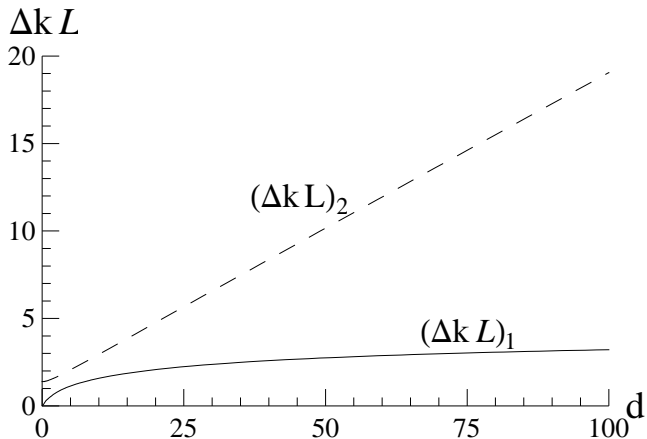}
\end{center}
\caption{If the two  metastable states are not degenerate, the efficiency of storage followed by backward retrieval will be lowered relative to the degenerate case, because the energy difference $\hbar\omega_{sg}$ introduces a momentum difference $\Delta k=\omega_{sg}/c$ between the quantum and classical fields.
As a function of $d$, the figure shows the momentum  $\Delta k L$, at which the optimal total efficiency of storage followed by backward retrieval falls to half of the $\Delta k L = 0$ value (dashed), and at which it is decreased to the optimal efficiency with forward retrieval (solid). \label{fig:modematchingKL}}
\end{figure}

We have, thus, demonstrated that we can optimize storage followed by backward retrieval for any given $d$ and $\Delta k$. We also recall that we have shown in Sec.~\ref{sec:freeadstret} that for $\Delta k = 0$ optimal storage followed by retrieval is accomplished with backward retrieval. However, as we increase $\Delta k L$, the optimal total efficiency with backward retrieval will drop down to the optimal total efficiency with forward retrieval at some value of $(\Delta k L)_1$. Increasing $\Delta k L$ further up to another value $(\Delta k L)_2$ will decrease the optimal total efficiency with backward retrieval to half of its $\Delta k L = 0$ value (and then further down to zero). Figure \ref{fig:modematchingKL} shows a plot of $(\Delta k L)_1$ (solid) and $(\Delta k L)_2$ (dashed) as a function of $d$. As noted above, without reoptimization, $(\Delta k L)_2$ would go as $\sqrt{d}$, but with optimization we see that it is linear in $d$, i.e., optimization makes the error less severe. $(\Delta k L)_1$ grows even slower than $\sqrt{d}$. This is not surprising because at $\Delta k L = 0$ optimal forward and optimal backward errors both fall off as $1/d$, except with different coefficients and, thus, eventually get very close to each other, so it takes a small $\Delta k L$ to make them equal. 

\begin{figure}[htb]
\begin{center}
\includegraphics[scale = 1]{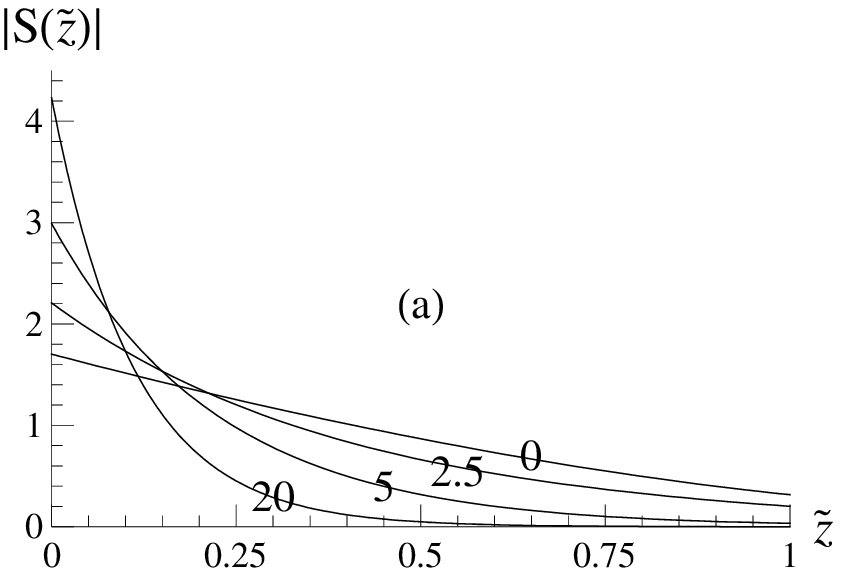}
\includegraphics[scale = 1]{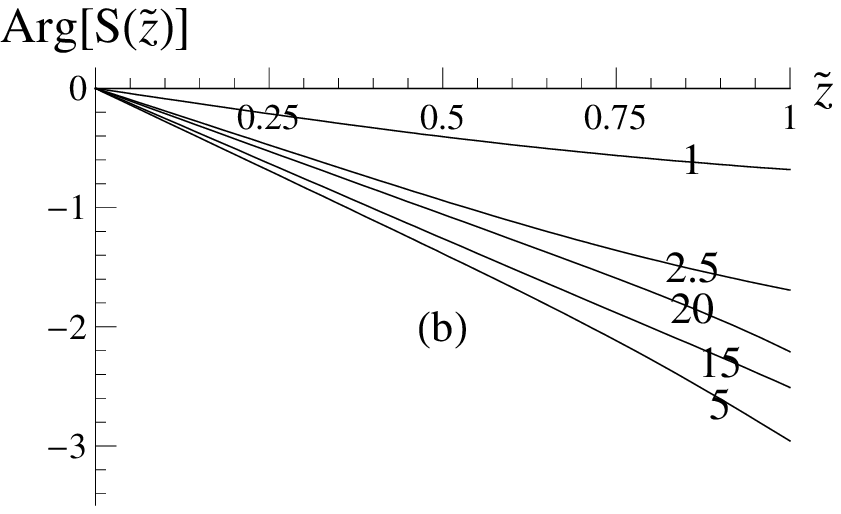}
\end{center}
\caption{(a) The magnitude and (b) the phase of the optimal mode for storage followed by backward retrieval at $d = 20$ for the indicated values of $\Delta k L$. The phase of the optimal mode at $\Delta k L = 0$ is $0$. The phase is plotted for the forward-propagation slowly varying operators as defined in Eq.~(\ref{slowops4}). \label{fig:modematchingmode}}
\end{figure}

In Figs.~\ref{fig:modematchingmode}(a) and \ref{fig:modematchingmode}(b), we show the magnitude $\left|S(\tilde z)\right|$ and the phase $\textrm{Arg}\left[S(\tilde z)\right]$, respectively, of the optimal mode (defined for the forward-propagating slowly varying operators as in Eq.~(\ref{slowops4})) at $d=20$ for different values of $\Delta k L$. As we increase $\Delta k L$, the optimal mode becomes concentrated more and more near the back end, i.e., it becomes favorable to effectively decrease the optical depth (i.e., decrease effective $L$) in order to decrease effective $\Delta k L$. The phase of the optimal mode is approximately linear, i.e., $S(\tilde z) \propto \exp(-i \tilde k_0 \tilde z)$ for some $\tilde k_0$. At $\Delta k L = 0$, $\tilde k_0 = 0$. Interestingly, instead of just growing from $0$ linearly with $\Delta k L$, $\tilde k_0$ first increases but then above $\Delta k L \sim 7.5$ starts decreasing again.

\section{Summary \label{sec:freesum}}

In conclusion, in this paper, we have presented a detailed analysis of the storage and retrieval of photons in homogeneously broadened $\Lambda$-type atomic media in free space and made a comparison to the cavity model described in paper I. From the investigation in the present paper emerges a new physical picture of the process of storage and retrieval in this system: first of all, the retrieval is essentially an interference effect where the emission from all the atoms interferes constructively in the forward direction. This constructive interference enhances the effective decay rate into the forward direction to $d \gamma$. The branching ratio between the desired forward radiation and the unwanted spontaneous emission is then simply given by the ratio between the various decay rates and is $\eta\sim d\gamma/(\gamma+d\gamma)\sim 1- 1/d$, irrespective of the method being used to drive the excitation out of the atoms. Secondly, the storage process is most conveniently viewed as the time reverse of retrieval. 

In the present paper, we have used this physical picture to derive the optimal strategy for storage and retrieval and the optimal efficiency that is independent of whether one works in the Raman, EIT, photon-echo, or any other intermediate regime. In particular, we showed how to achieve the optimal storage of any smooth input mode at any detuning of duration $T \gg 1/(d \gamma)$ (the adiabatic limit, including Raman and EIT) and of a particular class of resonant input modes of duration $T \sim 1/(d \gamma)$ (the fast or photon-echo limit). This analysis is extendable to other systems. In particular, in paper III, we consider the effects of inhomogeneous broadening on photon storage in $\Lambda$-type atomic media. Extensions to other systems, such as the double-$\Lambda$ system \cite{lvovsky06} or the tripod system \cite{zaremba07}, should also be possible.

We also suggested a novel time reversal based iterative procedure for optimizing quantum state mappings. Moreover, we showed that for the case of photon storage, this procedure is not only a convenient mathematical tool but is also a readily accessible experimental technique for finding optimal spin waves and optimal input-control pairs: one just has to be able to measure the output mode and to generate its time reverse. Following the present work, this procedure has recently been implemented experimentally with classical light \cite{novikova07}. We also expect this optimization procedure to be applicable to other systems used for light storage, such as tunable photonic crystals \cite{Yanik04}.

The presented optimization of the storage and retrieval processes leads to a substantial increase in the memory efficiency whenever reasonable synchronization between the input photon wave packet and the control pulse can be achieved. We, therefore, expect this work to be important in improving the efficiencies in current experiments, where optical depth is limited by various experimental imperfections such as a limited number of atoms in a trap \cite{kuzmich05}, competing four-wave mixing processes in a warm vapor cell \cite{eisaman05}, or inhomogeneous broadening in solid state samples \cite{kroll05}. 

\section{Acknowledgments}
We thank M.~Fleischhauer, M.~D.~Eisaman, E.~Polzik, J.~H.~M\"{u}ller, A.~Peng, J.~Nunn, I.~Novikova, D.~F.~Phillips, R.~L.~Walsworth, M.~Hohensee, M.~Klein, Y.~Xiao, N.\ Khaneja, A.~S.~Zibrov, P.~Walther, and A.~Nemiroski for fruitful discussions. This work was supported by the NSF, Danish Natural Science Research Council, DARPA, Harvard-MIT CUA, and Sloan and Packard Foundations.


\appendix
\section{Details of the Model and Derivation of the Equations of Motion \label{sec:appModelfree}}

In Sec.~\ref{sec:freespacemodel}, we presented a short introduction to the model and stated the equations of motion without derivation. In this Appendix, we provide the details of the model, as well as the derivation of the equations of motion (\ref{freeeqs2e})-(\ref{freeeqs2s}). Since the model and the assumptions made are very similar to those presented in the cavity case in paper I, we will often review some of them only briefly.

The electric field vector operator for the quantum field is given by \cite{loudon00}  
\begin{equation}
\mathbf{\hat E}_1(z) = \mathbf{\epsilon}_1 \left(\frac{\hbar \omega_1}{4 \pi c \epsilon_0 A}\right)^{1/2} \int d \omega \hat a_\omega e^{i \omega z/c} + h. c.,
\end{equation}
where $\textrm{h.c.}$ stands for Hermitian conjugate and where we have a continuum of annihilation operators $\hat a_\omega$ for the field modes of different frequencies $\omega$ that satisfy the commutation relation
\begin{equation}
\left[\hat a_\omega, \hat a^\dagger_{\omega'} \right] = \delta(\omega-\omega').
\end{equation}
By assumption, the field modes corresponding to $\hat a_\omega$ for different $\omega$ have the same transverse profile and are nonempty only around $\omega = \omega_1$. We have here assumed that the cross section $A$ of the beam is identical to the cross section of the ensemble. In typical experiments, the beam is smaller than the size of the ensemble, and in this case the relevant number of atoms $N$ should only be the number of atoms interacting with the beam. However, as we see from the final equations (\ref{freeeqs2e})-(\ref{freeeqs2s}), the only relevant quantity is the optical depth $d$, which does not depend on the area $A$, so that when everything is expressed in terms of $d$, the precise definition of $N$ and $A$ is irrelevant (see the end of this Appendix). 

The copropagating classical control field vector
\begin{equation}
\mathbf{E}_2(z,t) = \mathbf{\epsilon}_2 \eop_2(t-z/c)\cos(\omega_2(t-z/c))
\end{equation}
is a plane wave with polarization unit vector $\mathbf{\epsilon}_2$ and carrier frequency $\omega_2$ modulated by an envelope $\eop_2(t-z/c)$, which we assume to be propagating with group velocity equal to the speed of light $c$ since almost all the atoms are assumed to be in the ground state $|g\rangle$ and are, thus, unable to significantly alter the propagation of a strong classical field coupled to the $|s\rangle$-$|e\rangle$ transition. 

The Hamiltonian in Eqs.~(A3)-(A5) in paper I is then modified to
\begin{eqnarray} \label{freeH1} 
\hat H &=& \hat H_0 + \hat V,
\\ \label{freeH2}
\hat H_0 & = & \int d \omega \hbar \omega \hat a_\omega^\dagger \hat a_\omega + \sum^N_{i=1} \left(\hbar \omega_{se} \hat \sigma^i_{ss} + \hbar \omega_{ge} \hat \sigma^i_{ee}\right),
\\ \nonumber  
\hat V & = & - \hbar \sum^N_{i=1} \Big(\Omega (t-z_i/c) \hat \sigma^i_{es} e^{-i \omega_2 (t-z_i/c)}
\\ \label{freeH3}
& & + g \sqrt{\frac{L}{2 \pi c}} \int d \omega \hat a_\omega e^{i \omega z_i/c}\hat \sigma^i_{eg} + h.c\Big). 
\end{eqnarray}
Here $\hat \sigma^i_{\mu \nu} = |\mu\rangle_{i i}\langle\nu|$ is the internal state operator of the $i$th atom between states $|\mu\rangle$ and $|\nu \rangle$, $z_i$ is the position of the $i$th atom, $\mathbf{\hat d}$ is the dipole moment vector operator, $\Omega(t-z/c) = \langle e |(\mathbf{\hat d} \cdot \mathbf{\epsilon}_2)|s \rangle \eop_2 (t-z/c)/(2 \hbar)$ is the Rabi frequency of the classical field, and $g = \langle e |(\mathbf{\hat d} \cdot \mathbf{\epsilon}_1)|g \rangle  \sqrt{\frac{\omega_1}{2 \hbar  \epsilon_0 A L}}$ (assumed to be real for simplicity) is the coupling constant between the atoms and the quantized field mode, where we have chosen the length of the quantization volume to be identical to the ensemble length (this choice does not affect the results obtained below). We note that in order to avoid carrying extra factors of $2$ around, $\Omega$ is defined as half of the traditional definition of the Rabi frequency, so that a $\pi$ pulse, for example, takes time $\pi/(2 \Omega)$.

Since the position dependence along the ensemble matters, we divide our ensemble into thin slices along the length $L$ of the ensemble ($z=0$ to $z=L$) and introduce slowly varying operators
\begin{eqnarray} \label{slowops1}
\hat \sigma_{\mu \mu}(z,t) &=& \frac{1}{N_z} \sum_{i=1}^{N_z} \hat \sigma_{\mu \mu}^{i}(t),
\\
\hat \sigma_{es}(z,t) &=& \frac{1}{N_z} \sum_{i=1}^{N_z} \hat \sigma_{es}^{i}(t) e^{-i \omega_2 (t-z_i/c)},
\\
\hat \sigma_{eg}(z,t) &=& \frac{1}{N_z} \sum_{i=1}^{N_z} \hat \sigma_{eg}^{i}(t) e^{-i \omega_1 (t-z_i/c)},
\\ \label{slowops4}
\hat \sigma_{sg}(z,t) &=& \frac{1}{N_z} \sum_{i=1}^{N_z} \hat \sigma_{sg}^{i}(t) e^{-i (\omega_1-\omega_2) (t-z_i/c)},
\\ \label{slowops5}
\hat \eop(z,t) &=&  \sqrt{\frac{L}{2 \pi c}} e^{i \omega_1 (t-z/c)} \int d\omega \hat a_\omega(t) e^{i \omega z/c},
\end{eqnarray}
where sums are over all $N_z$ atoms in a slice of atoms positioned at $z$ that is thick enough to contain $N_z \gg 1$ atoms but thin enough that the resulting collective fields can be considered continuous. The normalization of $\hat \eop$ is chosen to ensure that it is dimensionless, which will be necessary to yield the simple dimensionless expressions in Eqs.~(\ref{freeeqs2e})-(\ref{freeeqs2s}). For these slowly varying operators, the effective Hamiltonian is
\begin{eqnarray} \nonumber \label{freeham}
\hat {\tilde H} &=& \int d \omega \hbar \omega \hat a_\omega^\dagger \hat a_\omega - \hbar \omega_1 \frac{1}{L} \int_0^L d z \eop^\dagger(z,t) \eop(z,t) 
\\ \nonumber
&&+ \int_0^L dz \hbar n(z) \bigg[ \Delta \hat \sigma_{ee}(z,t)- \Big(\Omega (t-z/c) \hat \sigma_{es}(z,t) 
\\ 
&& + g  \hat \eop(z,t) \hat \sigma_{eg}(z,t) + h. c.\Big)\bigg],
\end{eqnarray}
and the same-time commutation relations are
\begin{eqnarray} \nonumber
\left[\hat \sigma_{\mu \nu}(z,t), \hat \sigma_{\alpha \beta}(z',t)\right] \!\! &=& \!\! \frac{1}{n(z)} \left(\delta_{\nu \alpha} \hat \sigma_{\mu \beta}(z,t) \!-\! \delta_{\mu \beta} \hat \sigma_{\alpha \nu}(z,t)\right) 
\\ \label{freeatcom}
&&\times \delta(z-z'),
\\
\left[\hat \eop(z,t),\hat \eop^\dagger(z',t)\right] \!\!&=&\!\! L \delta(z-z').
\end{eqnarray}
Under the same assumptions as in the cavity case in paper I and defining $\hat P = \sqrt{N} \hat \sigma_{ge}$ and $\hat S = \sqrt{N} \hat \sigma_{gs}$, the Heisenberg equations of motion yield Eqs.~(\ref{freeeqse})-(\ref{freeeqss}), where, as in the cavity case in paper I, $\gamma$ may include extra dephasing in addition to radiative decay. We note that $\hat P$ and $\hat S$ are defined to be dimensionless in order to yield fully dimensionless Eqs.~(\ref{freeeqs2e})-(\ref{freeeqs2s}). The $\sqrt{N}$ in the definitions of $\hat P$ and $\hat S$ is required in order to have the final dimensionless equations depend on $g$, $N$, and $L$ only through the optical depth $d$. Similarly to the cavity case in paper I, from the generalized Einstein relations, the only nonzero noise correlations are \cite{hald01}
\begin{eqnarray}
\langle \hat F_P(z,t) \hat F^\dagger_P(z',t') \rangle &=& \frac{N}{n(z)} \delta(z-z') \delta(t-t'),
\\
 \langle \hat F_S(z,t) \hat F^\dagger_S(z',t') \rangle &=& \frac{N}{n(z)} \delta(z-z') \delta(t-t').
\end{eqnarray} 
Again the fact that normally ordered correlations are zero, as in the cavity case in paper I, means that the incoming noise is vacuum, which is precisely the reason why, as noted in Sec.~II in paper I, efficiency is the only number we need in order to fully characterize the mapping. The property of our system that guarantees that the incoming noise is vacuum is the absence of decay out of state $|g\rangle$ into states $|e\rangle$ and $|s\rangle$. We refer the reader to Appendix A of paper I for a detailed discussion of why this is a reasonable assumption in most experimental realizations.
 
We will now show how our field and atomic operators can be expanded in terms of modes, which is necessary in order to obtain and interpret the final complex number equations (\ref{freeeqs2e})-(\ref{freeeqs2s}). Under the assumption that almost all atoms are in the ground state at all times, commutation relations (\ref{freeatcom}) imply
\begin{eqnarray} \label{freeScom}
\left[\hat S(z,t),\hat S^\dagger(z',t)\right] = \frac{N}{n(z)} \delta(z-z'),
\\
\left[\hat P(z,t),\hat P^\dagger(z',t)\right] = \frac{N}{n(z)} \delta(z-z'). 
\end{eqnarray}
Equation (\ref{freeScom}) allows us to expand $\hat S(z,t)$ in terms of any basis set of spatial modes $\left\{g_\alpha(z)\right\}$ satisfying the orthonormality relation $\int_0^L d z g^*_\alpha(z) g_\beta(z) = \delta_{\alpha \beta}$ and the completeness relation $\sum_\alpha g^*_\alpha(z) g_\alpha(z') = \delta(z-z')$ as
\begin{equation}
\hat S(z,t) = \sqrt{\frac{N}{n(z)}} \sum_\alpha g_\alpha(z) \hat c_\alpha(t),
\end{equation}
where the annihilation operators $\left\{\hat c_\alpha\right\}$ for the spin-wave modes satisfy
\begin{equation}
\left[\hat c_\alpha(t),\hat c_\beta^\dagger(t)\right] = \delta_{\alpha \beta}.
\end{equation}

For the freely propagating input field $\hat \eop_\textrm{in}(t) = \hat \eop(0,t)$ and output field $\hat \eop_\textrm{out}(t) = \hat \eop(L,t)$ we have the following commutation relations:
\begin{eqnarray}
\left[\hat \eop_\textrm{in}(t),\hat \eop^\dagger_\textrm{in}(t')\right] &=& \frac{L}{c} \delta(t-t'),
\\
\left[\hat \eop_\textrm{out}(t),\hat \eop^\dagger_\textrm{out}(t')\right] &=& \frac{L}{c} \delta(t-t'),
\end{eqnarray} 
which differ from their cavity case counterparts in Eq.~(A19) of paper I only in normalization. These commutation relations allow
us to expand, as in the cavity case in paper I, the input and the output field in terms of any basis set of field (envelope) modes $\left\{h_\alpha(t) \right\}$ defined for $t \in [0,\infty)$, satisfying the orthonormality relation $\int_0^\infty d t h^*_\alpha(t) h_\beta(t) = \delta_{\alpha \beta}$ and the completeness relation $\sum_\alpha h^*_\alpha(t) h_\alpha(t') = \delta(t-t')$, as
\begin{eqnarray}
\hat \eop_\textrm{in}(t) = \sqrt{\frac{L}{c}}\sum_\alpha h_\alpha(t) \hat a_\alpha,
\\
\hat \eop_\textrm{out}(t) = \sqrt{\frac{L}{c}} \sum_\alpha h_\alpha(t) \hat b_\alpha,
\end{eqnarray}
where annihilation operators $\left\{\hat a_\alpha \right\}$ and $\left\{\hat b_\alpha \right\}$ for the input and the output photon modes, respectively, satisfy the usual bosonic commutation relations (see Eq.\ (A22) in paper I).

All atoms are initially pumped into the ground state, i.e., no $\hat P$ or $\hat S$ excitations are present in the atoms. We also assume that the only input field excitations initially present are in the quantum field mode with annihilation operator $\hat a_0$ corresponding to an envelope shape $h_0(t)$ nonzero on $[0,T]$. Precisely as in the cavity case in paper I, the only parts of the operators that will contribute to the efficiency will be the parts proportional to $\hat a_0$. We can therefore reduce our problem to complex number equations. These equations and the corresponding initial and boundary conditions are given in Sec.~\ref{sec:freespacemodel}. To get back the nonvacuum part of the original operator from its complex number counterpart, one can just multiply the complex number version by $\hat a_0$.
 
We conclude this Appendix with a verification that $d = g^2 N L/(\gamma c)$ is independent of the size of the beam and, for a given transition, depends only on the density of the atoms and the length of the ensemble. This can be seen directly by inserting the definition of $g$ into $d$ and defining the atomic number density $\rho(z) = n(z)/A$ (which, by assumption, is uniform in the direction transverse to the propagation direction). The expression for $d$ then becomes $d = |\langle e |(\mathbf{\hat d} \cdot \mathbf{\epsilon}_1)|g \rangle|^2  \omega_1\int \rho(z)dz /(2 \hbar  \epsilon_0\gamma c)$. So it is indeed independent of the size of the beam and, for a given transition, only depends on the density and length of the ensemble.

\section{Position Dependence of Loss \label{sec:position}}

We have shown in Sec.~\ref{sec:freeret} that, provided the retrieval control pulse is long and/or powerful enough to leave no atomic excitations (i.e., the retrieval is complete), the retrieval efficiency $\eta_\textrm{r}$ depends only on the optical depth $d$ and the spin wave $S(\tilde z)$ but not on the detuning $\tilde \Delta$ and the control field envelope $\tilde \Omega(\tilde t)$. In this Appendix, we show that for complete retrieval, not only the total efficiency but also the distribution of spontaneous emission loss (or more precisely loss due to polarization decay $\gamma$) as a function of position is independent of the control and the detuning. 

Equations of motion (\ref{freeeqs2e})-(\ref{freeeqs2s}) imply that
\begin{equation}\label{appeq1}
\partial_{\tilde z} |\eop(\tilde z, \tilde t)|^2
+ \partial_{\tilde t} |P(\tilde z, \tilde t)|^2
+ \partial_{\tilde t} |S(\tilde z, \tilde t)|^2
= -2 |P(\tilde z, \tilde t)|^2.
\end{equation}
Integrating both sides with respect to $\tilde z$ from $0$ to $1$ and with respect to $\tilde t$ from $\tilde T_{\textrm{r}}$ to $\infty$, using the initial conditions $S(\tilde z, \tilde T_{\textrm{r}}) = S(\tilde z)$ (where $\int_0^1 d \tilde z |S(\tilde z)|^2 = 1$) and $P(\tilde z, \tilde T_{\textrm{r}}) = 0$, the boundary condition $\eop(0, \tilde t) = 0$, and the complete retrieval condition $S(\tilde z,\infty) = P(\tilde z, \infty) = 0$, we find, using Eq.~(\ref{etar}), that 
\begin{equation}
\eta_\textrm{r} = 1 - \int_0^1\!\! d \tilde z \; l(\tilde z),
\end{equation}
where the position-dependent loss per unit length is
\begin{equation}
l(\tilde z) = 2 \int_0^\infty\!\! d \tilde t \left| P(\tilde z, \tilde t)\right|^2.
\end{equation}
Computing $l(\tilde z)$, we find
\begin{eqnarray}
l(\tilde z) &=& 2 \mathcal{L}^{-1} \left\{\int_{\tilde T_{\textrm{r}}}^\infty d \tilde t P(u,\tilde t) \left[P(u'^*,\tilde t)\right]^*\right\}_{u,u' \rightarrow \tilde z}
\nonumber \\
&=& \mathcal{L}^{-1} \left\{\frac{2}{2 + \frac{d}{u} + \frac{d}{u'}}  S(u) \left[S(u'^*)\right]^*\right\}_{u,u' \rightarrow \tilde z},
\end{eqnarray}
where $\mathcal{L}^{-1}$ with subscript $u,u' \rightarrow \tilde z$ means that inverse Laplace transforms are taken with respect to $u$ and $u'$ and are both evaluated at $\tilde z$. In the last equality, Eq.~(\ref{ddtPPstar}) and the conditions at $\tilde t=\tilde T_{\textrm{r}}$ and $\tilde t=\infty$ were used. Therefore, we see that $l(\tilde z)$ is independent of the detuning and the control. Moreover, the inverse Laplace transforms $\mathcal{L}^{-1}$ can be taken analytically to give 
\begin{eqnarray} \label{Lofz}
l(\tilde z)\!\! &=& \!\! |S(\tilde z)|^2 - \textrm{Re}\left[S(\tilde z) d \int_0^{\tilde z} d \tilde z' S^*(\tilde z- \tilde z') e^{- \frac{d \tilde z'}{2}}\right]
\nonumber \\
&+& \!\! \int_0^{\tilde z} \! d \tilde z' \int_0^{\tilde z} \! d \tilde z'' S(\tilde z- \tilde z') S^*(\tilde z-\tilde z'') \frac{d^2}{4} e^{- \frac{d}{2}(\tilde z' +\tilde z'')} 
\nonumber \\
&\times &\!\! \left[2 I_0\left(d \sqrt{\tilde z'\tilde z''}\right) - \frac{\tilde z' +\tilde z''}{\sqrt{\tilde z'\tilde  z''}} I_1\left(d \sqrt{\tilde z' \tilde z''}\right)\right].  
\end{eqnarray}

\section{Implementation of the Inverse Propagator using Time Reversal \label{sec:apptimerev}}

In Sec.~\ref{sec:timerev}, we exploited the fact that time reversal could be used to realize the inverse evolution $\hat U^{-1}[T,0; \Omega(t)]$. In this Appendix, we would like to explain carefully what we mean by the time reversal operator $\hat{\mathcal{T}}$ and to prove Eq.~(\ref{inverseprop}). 

To define the time reversal operator $\hat{\mathcal{T}}$, we first choose a basis for our single-excitation Hilbert space consisting of $\{\hat \sigma^i_{sg} |\textrm{ground}\rangle\}$, $\{\hat \sigma^i_{eg} |\textrm{ground}\rangle\}$, and $\{\hat a^\dagger_z |\textrm{ground}\rangle\}$, where $|\textrm{ground}\rangle$ is the state with no photons and no atomic excitations (i.e., all atoms in the ground state), $i$ runs over all atoms, $z$ runs over all positions, and $\hat a^\dagger_z = (2 \pi c)^{-1/2} \int d \omega \exp\left(-i \omega z/c\right) \hat a^\dagger_\omega$. We then define the time reversal operator $\hat{\mathcal{T}}$ (equivalent to the complex conjugation operator $K$ in Ref.~\cite{sakurai94}) as follows: $\hat{\mathcal{T}}|\psi(t)\rangle$ means taking the complex conjugates of the expansion coefficients of a state $|\psi(t)\rangle$ in the above basis, while $\hat{\mathcal{T}} \hat O \hat{\mathcal{T}}$ means taking complex conjugates of the matrix elements of the operator $\hat O$ when $\hat O$ is written in the above basis (we will, thus, write $\hat{\mathcal{T}} \hat O \hat{\mathcal{T}} = \hat O^*$). For example, this definition implies that in addition to complex conjugating the envelope of the photon, time reversal flips the photon momentum: $\hat{\mathcal{T}} a_\omega \hat{\mathcal{T}} = \hat a_{-\omega}$ and $\hat{\mathcal{T}} \hat a_\omega |\textrm{ground}\rangle = \hat a_{-\omega} |\textrm{ground}\rangle$. Some of the properties of $\hat{\mathcal{T}}$ are $\hat{\mathcal{T}}^2 = \hat{\textbf{1}}$ and $|\langle \psi_1|\hat{\mathcal{T}} |\psi_2\rangle | = |\langle \hat{\mathcal{T}} \psi_1| \psi_2 \rangle|$. 

We now turn to the proof of Eq.~(\ref{inverseprop}). We start by noting that
\begin{eqnarray} \label{appinverseprop}
 \hat U^{-1}[T,0;\Omega(t)]  &=& \hat U[0,T;\Omega(t)] = \hat{\mathcal{T}} \hat{\mathcal{T}} \hat U[0,T;\Omega(t)] \hat{\mathcal{T}} \hat{\mathcal{T}} 
\nonumber \\
&=& \hat{\mathcal{T}} \hat U^*[0,T;\Omega(t)] \hat{\mathcal{T}}
\end{eqnarray}
and, therefore, using Eq.~(\ref{etaUinv}),
\begin{equation} \label{etaUstar}
\eta=|\langle b |\hat U[T,0;\Omega(t)]|a\rangle|^2=|\langle a |\hat{\mathcal{T}} \hat U^*[0,T;\Omega(t)] \hat{\mathcal{T}}|b\rangle|^2.
\end{equation}
To evaluate $\hat U^*[0,T;\Omega(t)]$ and to find a way to implement it physically, let us first consider the simplest case, where the Hamiltonian responsible for the evolution is independent of time and respects time reversal symmetry $\hat{\mathcal{T}} \hat H \hat{\mathcal{T}}= \hat H$. This is equivalent to 
$\hat H^*=\hat H$. In this case, the evolution operator is
given by $\hat U[T,0]=\exp(-i \hat H T/\hbar)$, and, therefore, $\hat U^{*}[0,T] = \hat U[T,0]$. So if the Hamiltonian obeys time reversal symmetry, one can physically implement $\hat U^*[0,T]$ simply by evolving the system for a time $T$. Applied to Eq.~(\ref{etaUstar}), this would mean that the probability to go from $|a\rangle$ to $|b\rangle$ due to unitary evolution $\hat U$ is the same as the probability to make the transition from $\hat{\mathcal{T}}|b\rangle$ to $\hat{\mathcal{T}}|a\rangle$ due to this evolution. In other words, according to time reversal, if our Hamiltonian obeyed time reversal symmetry, we would be able to map the time reverse $\eop^*(T-t)$ of the output mode onto the spin wave $S^*$ with the overlap storage efficiency equal to the retrieval efficiency from the spin wave $S$. 

In general, the interaction does not obey time reversal symmetry because of the classical control, which may depend on time and may be complex. To extend the discussion to this situation, we shall use the equation  of motion for the propagator 
\begin{eqnarray}
\label{propagatoreq1}
i\hbar \frac{d \hat U[\tau_1,\tau_2;\Omega(t)]}{d\tau_1}=\hat H[\tau_1; \Omega(t)] \hat U[\tau_1,\tau_2;\Omega(t)],
\end{eqnarray}
where we have highlighted the dependence of the Hamiltonian on the control field by including $\Omega(t)$ as an argument of $\hat H$. By taking the complex conjugate of Eq.~(\ref{propagatoreq1}), we can turn it into the equation for the time-reversed inverse propagator $\hat{\mathcal{T}}\hat U^{-1}[\tau_2,\tau_1;\Omega(t)]\hat{\mathcal{T}}=\hat U^*[\tau_1,\tau_2;\Omega(t)]$:
\begin{equation}\label{propagatoreq2}
-i\hbar \frac{d \hat U^*[\tau_1,\tau_2;\Omega(t)]}{d\tau_1}=\hat H^*[\tau_1; \Omega(t)]\hat U^*[\tau_1,\tau_2;\Omega(t)].
\end{equation}
Note that we have not made any assumptions about time reversal symmetry being a symmetry for the system.  

In order to implement the evolution in the last expression of Eq.~(\ref{etaUstar}), we need to realize a time evolution $\hat U[T,0;\Omega'(t)]$ such that $\hat U[T,0;\Omega'(t)]=\hat U^*[0,T,\Omega(t)]$. To do this, we consider operators $\hat U[\tau,0;\Omega'(t)]$ and $\hat U^*[T-\tau,T,\Omega(t)]$ whose equality at time $\tau = T$ will imply the desired evolution. The equation of motion for these operators are, using Eqs.~(\ref{propagatoreq1}) and (\ref{propagatoreq2}),
\begin{eqnarray}
i\hbar \frac{d \hat U^*[T\!-\!\tau,T;\Omega(t)]}{d\tau}\!\!&=&\!\!\hat H^*[T\!-\!\tau;\Omega(t)]\hat U^*[T\!-\!\tau,T;\Omega(t)],
\nonumber \\
i\hbar \frac{d \hat U[\tau,0;\Omega'(t)]}{d\tau}\!\!&=&\!\!\hat H[\tau; \Omega'(t)]\hat U[\tau,0;\Omega'(t)].
\end{eqnarray}
Since the two operators are the same at time $\tau=0$, $\hat U[0,0,\Omega'(t)]=\hat U^*[T,T,\Omega(t)]=\hat{\textbf{1}}$, the two operators will be identical (and, in particular, equal at $\tau = T$) if they obey the same differential equation, which is the case if 
\begin{equation} \label{HH}
\hat H[t, \Omega'(t)]=\hat H^*[T-t; \Omega(t)].
\end{equation}
We would like to evaluate the right-hand side of Eq.~(\ref{HH}). By inspecting the Hamiltonian in Eqs.~(\ref{freeH1})-(\ref{freeH3}), we see that the only non-trivial parts are the classical control field (including the carrier $\omega_2$) and the exponential in the interaction with the quantum field (the part of Eq.~(\ref{freeH3}) containing $g$). According to the right hand side of Eq.~(\ref{HH}), we would like to apply the complex conjugation to $\hat H$ and evaluate it at time $T-t$. The interaction with the quantum field  is actually invariant not only under the change of time but also under complex conjugation since applying time reversal to $\hat a_\omega e^{i \omega z_i/c}$ simply changes $\omega$ to $-\omega$, which can be changed back to $\omega$ by flipping the sign of the integration variable. The application of complex conjugation and the change of time to the interaction with the classical field is equivalent to using the time-reversed control field envelope $\Omega'(t) = \Omega^*(T-t)$ and a carrier wave vector propagating in the opposite direction. Combining this result with Eq.~(\ref{appinverseprop}), we arrive at Eq.~(\ref{inverseprop}). This means that by using the time-reversed control field $\Omega^*(T-t)$, we can map $\hat{\mathcal{T}}|b\rangle$ onto  $\hat{\mathcal{T}}|a\rangle$ with the probability equal to the probability that $|a\rangle$ goes to $|b\rangle$ using $\Omega(t)$. 

\section{Proof of Convergence of Optimization Iterations to the Optimum \label{sec:apptimerev2}}

In Sec.~\ref{sec:timerev2}, we omitted the proof that  iterative application of $\hat{\mathcal{N}} \hat P_A \hat U^{-1} \hat P_B \hat U$ to a unit vector $|a\rangle \in A$   converges to $|a_\textrm{max}\rangle$ (unless $\langle a|a_\textrm{max}\rangle = 0$) and that $\hat{\mathcal{N}} \hat P_B \hat U |a_\textrm{max}\rangle$ optimizes $\hat U^{-1}$ as a map from $B$ to $A$. In this appendix, we present this proof.

For any unit vector $|a\rangle$ in the subspace $A$ of ``initial" states, the efficiency is defined as $\eta= |\hat P_B \hat U |a\rangle|^2$, where $\hat P_B$ denotes the projection on the subspace $B$ of ``final" states. We are looking for $|a_\textrm{max}\rangle \in A$ that gives the maximum efficiency $\eta_\textrm{max} = |\hat P_B \hat U |a_\textrm{max}\rangle|^2$. By appropriately adjusting the phases, we can write $\sqrt{\eta_\textrm{max}} = \langle b_\textrm{max}|\hat U|a_\textrm{max} \rangle$ for some unit vector $|b_\textrm{max} \rangle =\hat{\mathcal{N}} \hat P_B \hat U|a_{\textrm{max}}\rangle \in B$. From the unitarity of $\hat U$ (see, for example, Eq.~(\ref{etaUinv})), it follows that $|b_\textrm{max} \rangle$ optimizes $\hat U^{-1}$ as a mapping from $B$ to $A$. We will prove two claims, from which the desired convergence result will follow immediately. Claim $(1)$: if $|a_1 \rangle$ is orthogonal to $|a_\textrm{max}\rangle$, $\hat U |a_1 \rangle$ is orthogonal to $|b_\textrm{max}\rangle$. Proof: suppose  $\langle b_\textrm{max}|\hat U|a_1 \rangle = \beta \neq 0$, then defining $|\tilde a \rangle  = (\sqrt{\eta_\textrm{max}} |a_\textrm{max}\rangle + \beta^* |a_1\rangle)/\sqrt{\eta_\textrm{max} + |\beta|^2}$, we have $\langle b_\textrm{max}|\hat U|\tilde a\rangle = \sqrt{\eta_\textrm{max} +|\beta|^2} > \sqrt{\eta_\textrm{max}}$, which contradicts the fact that $|a_\textrm{max}\rangle$ was optimal. A similar proof can be given for claim $(2)$: if $|b_1 \rangle$ is orthogonal to $|b_\textrm{max}\rangle$, $\hat U^{-1} |b_1 \rangle$ is orthogonal to $|a_\textrm{max}\rangle$. From these two claims it immediately follows that if we start with a state orthogonal to $|a_\textrm{max}\rangle$, we will never approach $|a_\textrm{max}\rangle$. On the other hand, we will show now that if we start with $|a \rangle = \alpha |a_\textrm{max}\rangle + \sqrt{1-|\alpha|^2} |a_1\rangle$ (for some unit $|a_1\rangle$ orthogonal to $|a_\textrm{max}\rangle$ and for some $\alpha \neq 0$), we will indeed approach $|a_\textrm{max}\rangle$. We have $\hat P_B \hat U |a\rangle = \alpha \langle b_\textrm{max}|\hat U|a_\textrm{max}\rangle |b_\textrm{max}\rangle + \sqrt{1-|\alpha|^2} \langle b_1|\hat U|a_1\rangle |b_1\rangle$ for some unit $|b_1\rangle \!\in\! B$. By claim $(1)$, the two parts of $\hat P_B \hat U |a\rangle$ are orthogonal and, since $|a_\textrm{max}\rangle$ is optimal, $|\langle b_\textrm{max}|\hat U|a_\textrm{max}\rangle| > |\langle b_1|\hat U|a_1\rangle|$. Thus, the fraction of $|b_\textrm{max}\rangle$ in $\hat P_B \hat U |a\rangle$ is greater than the fraction of $|a_\textrm{max}\rangle$ in $|a\rangle$. After the application of $\hat P_A \hat U^{-1}$ and during subsequent iterations, the optimal fraction will similarly grow. This shows that we will indeed reach the optimum, unless we start with something orthogonal to it.

\section{Shaping the Control Field for the Optimal Adiabatic Storage \label{sec:appstor}}

In this Appendix, we use Eq.~(\ref{freeads}) to find the control field for the storage of any given mode $\eop_\textrm{in}(\tilde t)$ into any given decayless spin-wave mode $s(\tilde z)$. We then verify that the optimal storage control found through this procedure using the optimal decayless spin-wave mode $s(\tilde z)$ gives storage into $\tilde S_d(1-\tilde z)$, the optimal mode for backward retrieval, with efficiency equal to the optimal retrieval efficiency $\eta^\textrm{max}_\textrm{r}$. We also verify that this control field is the time-reverse of the control field that retrieves the optimal spin-wave mode into $\eop^*_\textrm{in}(\tilde T-\tilde t)$, that is, the time-reverse of the input mode.

In order to solve for $\Omega(\tilde t)$ from Eq.~(\ref{freeads}), we note that $q(\tilde z,\tilde t)$ satisfies
\begin{eqnarray} \label{qzeq2}
\int_0^\infty d \tilde z q(\tilde z,\tilde t) q^*(\tilde z,\tilde t') = \delta(\tilde t-\tilde t'),
\\ \label{qteq2}
\int_0^T d \tilde t q(\tilde z,\tilde t) q^*(\tilde z',\tilde t) = \delta(\tilde z-\tilde z'),
\end{eqnarray}
where we have used the identity
\begin{equation}
\int_0^\infty d x J_0(a x) J_0(b x) x = \frac{1}{a} \delta(a-b), 
\end{equation}
and where Eq.~(\ref{qteq2}) requires $h(0,\tilde T) = \infty$ (we will discuss below that this requirement can be relaxed without significant loss in efficiency). Using Eq.~(\ref{qzeq2}), we see that $\int_0^\infty d \tilde z |s(\tilde z)|^2 = 1$, as expected from unitarity (since we neglect both the leakage and the decay rate $\gamma$ in Eq.~(\ref{freeads}), the transformation between $\eop_\textrm{in}(\tilde t)$ and $s(\tilde z)$ is unitary). Using Eq.~(\ref{qzeq2}), we can also invert Eq.~(\ref{freeads}) to get 
\begin{equation} \label{freeade}
\eop_\textrm{in}(\tilde t) = \int_0^\infty d \tilde z q^*(\tilde z,\tilde t) s(\tilde z).
\end{equation}
Clearly, Eqs.~(\ref{freeads}) and (\ref{freeade}) establish a 1-to-1 correspondence, for a given $\tilde \Omega(\tilde t)$, between input modes $\eop_\textrm{in}(\tilde t)$ and decayless modes $s(\tilde z)$. This 1-to-1 correspondence is the demonstration of the unitarity of the map defined by Eq.~(\ref{freeads}). For the purposes of shaping the control field, it is crucial that Eqs.~(\ref{freeads}) and (\ref{freeade}) also establish a $1$-to-$1$ correspondence, for a given $\eop_\textrm{in}(\tilde t)$, between controls $\tilde \Omega(\tilde t)$ (satisfying $h(0,\tilde T) = \infty$ and nonzero whenever $\eop_\textrm{in}(\tilde t)$ is nonzero) and normalized decayless propagation modes $s(\tilde z)$. In particular, Eq.~(\ref{freeads}) itself allows to determine $s(\tilde z)$ from $\tilde \Omega(\tilde t)$ and $\eop_\textrm{in}(\tilde t)$. To solve for $\tilde \Omega(t)$ given $s(\tilde z)$ and $\eop_\textrm{in}(t)$, we integrate, as in Sec.~\ref{sec:retshaping}, from $0$ to $\tilde t$ the norm squared of both sides of Eq.~(\ref{freeade}) and change the integration variable from $\tilde t'$ to $h'= h(\tilde t',\tilde T)$ on the right-hand side to obtain
\begin{eqnarray}\label{hequation2}
\!\!\!\!\!\!\!\!\!\!\!\!\!\int_0^{\tilde t} d \tilde t' |\eop_\textrm{in}(\tilde t')|^2 & = & \int_{h(\tilde t,\tilde T)}^{h(0,\tilde T)} d h' \frac{d}{\tilde \Delta^2} 
\nonumber \\
&& \!\!\!\!\!\!\!\!\!\!\!\!\! \times  \left|\int_0^\infty d \tilde z  e^{-i \frac{d \tilde z}{\tilde \Delta}} J_0\left(2 \sqrt{\frac{h' d \tilde z}{\tilde \Delta^2}}\right)  s(\tilde z)\right|^2.
\end{eqnarray}
Using $h(\tilde T,\tilde T) = 0$ and the normalization of $s(\tilde z)$ and $\eop_\textrm{in}(\tilde t)$, the evaluation of Eq.~(\ref{hequation2}) at $\tilde t = \tilde T$, implies that $h(0,\tilde T) = \infty$ (unless the expression inside the absolute value sign vanishes for all $h'$ greater than some finite value). This divergence is, however, just a mathematical convenience: truncating $|\Omega(t)|$ does not significantly affect the efficiency, as we discuss in Secs.~\ref{sec:retshaping}, \ref{sec:freeadst}, and \ref{sec:freeadcond}. We also give in Sec.~\ref{sec:freeadst} approximate expressions for how big the optimal $|\Omega|$ is in the Raman and resonant limits. Replacing $h(0,\tilde T)$ with $\infty$, we use Eq.~(\ref{hequation2}) to solve numerically for $h(\tilde t,\tilde T)$, exactly as in Sec.~\ref{sec:retshaping}. $|\tilde \Omega(\tilde t)|$ is then deduced by taking the square root of the negative of the derivative of $h(\tilde t,\tilde T)$. The phase of $\tilde \Omega$ is found by inserting $|\tilde \Omega|$ into Eq.~(\ref{freeade}) is given by
\begin{eqnarray}\label{freestorphase}
\!\!\!\!\!\!\!\!\!\!\!\!\! \textrm{Arg}\left[\tilde \Omega(\tilde t)\right] &=& \frac{\pi}{2} + \textrm{Arg}\left[\eop_\textrm{in}(\tilde t)\right] + \frac{h(\tilde t,\tilde T)}{\tilde \Delta} 
\nonumber \\ 
&&\!\!\!\!\!\!\!\!\!\!\!\!\!\!\!\!\!\!\!\!\!\!\!\! 
+ \textrm{Arg}\left[ \int_0^\infty d \tilde z e^{i \frac{d \tilde z}{\tilde \Delta}}  J_0\left(2 \sqrt{\frac{h(\tilde t,\tilde T) d \tilde z}{\tilde \Delta^2}}\right) s^*(\tilde z)\right].  
\end{eqnarray}
The various terms in Eq.~(\ref{freestorphase}) can be interpreted in a way similar to the terms in the phase of the retrieval control in Eq.~(\ref{freeretphase}) in Sec.~\ref{sec:retshaping}. The only minor difference in the interpretation is that in the limit $\tilde \Delta \rightarrow 0$, the third term seems to diverge. However, one can check that in this limit the last term cancels (up to a constant) with the third term to ensure that the phase of the optimal control is still given solely by the phase of the desired output, as expected for the resonant limit. Finally, we note that the same remarks as at the end of Sec.~\ref{sec:retshaping} regarding the ability to truncate the divergences of $|\tilde \Omega(\tilde t)|$ without significant loss in efficiency apply.

Using Eq.~(\ref{adddecay}), we show in Sec.~\ref{sec:freeadst} how to find the optimal decayless mode $s(z)$, which should then be used in Eq.~(\ref{freeads}) to shape the optimal control (using Eqs.~(\ref{hequation2}) and (\ref{freestorphase})). Having derived the optimal storage control in this way, we can now explicitly verify the results obtained from the time reversal reasoning. Two of these results are that the mode $S(\tilde z,\tilde T)$ used in optimal storage is just the optimal mode for backward retrieval and that the optimal storage efficiency and the optimal retrieval efficiency are equal. One can explicitly verify these statements by checking that the application of Eq.~(\ref{adddecay}) to the iteration used to find the optimal $s(\tilde z)$ gives the iteration used to find the optimal backward retrieval mode, i.e.,
\begin{equation}
S_2(\tilde z) = \int_0^1 d \tilde z' k_\textrm{r}(\tilde z,\tilde z') S_1(\tilde z'),
\end{equation}
where $k_\textrm{r}(\tilde z,\tilde z')$ is defined in Eq.~(\ref{retkernel}). Thus,  $S(\tilde z,\tilde T)$ that we compute from the optimal $s(\tilde z)$ using Eq.~(\ref{adddecay}) is indeed the optimal mode for backward retrieval, and the optimal storage efficiency is indeed equal to the optimal retrieval efficiency $\eta^\textrm{max}_\textrm{r}$:
\begin{equation} \label{SSd}
S(\tilde z,\tilde T) = \sqrt{\eta^\textrm{max}_\textrm{r}} \tilde S_d(1-\tilde z).
\end{equation} 

Another consequence of time reversal is the fact that the optimal storage control for a given input mode is the time reverse of the control that gives optimal backward retrieval into the time reverse of that input mode. One can verify this directly by comparing the expressions of the magnitude and phase of the two controls. However, a simpler approach is to consider $\eop_\textrm{out}(\tilde t)$ given by Eq.~(\ref{freeadeout}) for the case of retrieval with a certain control $\tilde \Omega(\tilde t)$ from the optimal mode, i.e., $S(\tilde z) = \tilde S_d(\tilde z)$. We then use Eq.~(\ref{freeadS}) to compute the spin wave $S_2(\tilde z)$ that results from storing $\eop^*_\textrm{out}(\tilde T - \tilde t)$ with $\tilde \Omega^*(\tilde T - \tilde t)$. In the limit $\tilde T \rightarrow \infty$ (to make sure that we fully retrieve $\eop_\textrm{out}$ before sending its time reverse back in), we can take the time integral explicitly to find 
\begin{equation}
S_2(\tilde z) = \int_0^1 d \tilde z' k_\textrm{r}(\tilde z,\tilde z') \tilde S_d(1-\tilde z') = \eta^\textrm{max}_\textrm{r} \tilde S_d(1-\tilde z), 
\end{equation}
where in the last step we used the definition of $\tilde S_d(1-\tilde z)$ as the eigenvector of $k_\textrm{r}$ with the largest eigenvalue  $\eta^\textrm{max}_\textrm{r}$ equal to the optimal retrieval efficiency. Thus, the total efficiency of optimal retrieval followed by time-reversed storage is $(\eta^\textrm{max}_\textrm{r})^2$. So we have shown explicitly that the time reverse of optimal retrieval gives storage into $\tilde S_d$ with the maximum efficiency $\eta^\textrm{max}_\textrm{r}$, confirming what we have shown in Sec.~\ref{sec:timerev} based on general time reversal arguments. Since we have shown in this Appendix that the optimal control field is unique, we have therefore confirmed that the control that optimally stores a given input and the control that optimal retrieves into the time reverse of that input are time-reverses of each other.

\end{document}